\documentclass[twoside,preprint]{article}
\usepackage[accepted]{aistats2020}

\usepackage{graphicx}
\usepackage{dsfont}
\usepackage{booktabs} 

\usepackage{amsmath, amssymb, bbm, amsthm}
\usepackage{nameref}
\usepackage{subcaption}
\usepackage{mdframed}
\usepackage{url} 
\usepackage[round]{natbib}

\usepackage{algpseudocode}

\usepackage{tikz} 
\usetikzlibrary{positioning,shapes.geometric} 
\usetikzlibrary{decorations.pathreplacing,angles,quotes}
\usepackage{enumitem}

\newcommand{\argmax}{\mathrm{argmax}}

\newcommand{\nk}{n^{(k)}}

\renewcommand{\P}{\mathbb{P}}
\newcommand{\E}{\mathbb{E}}
\newcommand{\Ind}{\mathbbm{1}}
\newcommand{\bbeta}{\boldsymbol{\beta}}
\newcommand{\excompas}{\textbf{Example: COMPAS.}}

\usepackage{mathtools}
\DeclarePairedDelimiter{\ceil}{\lceil}{\rceil}

\newtheorem{theorem}{Theorem}[section]
\newtheorem{corollary}{Corollary}[theorem]

\newtheorem{proposition}[theorem]{Proposition}
\newtheorem{assumption}{Assumption}

\graphicspath{{./figures/}{./figures_old/}}

\usepackage{hyperref}

\newcommand{\Pb}{\mathbb{P}}
\def\ind{\perp\!\!\!\perp}

\begin{document}


\twocolumn[

\aistatstitle{Fairness Evaluation in Presence of Biased Noisy Labels}

\aistatsauthor{ Riccardo Fogliato \And Max G'Sell \And  Alexandra Chouldechova }


\aistatsaddress{ Carnegie Mellon University \\ Partnership on AI \And  Carnegie Mellon University \And Carnegie Mellon University \\ Partnership on AI } ]

\begin{abstract}
  Risk assessment tools are widely used around the country to inform decision making within the criminal justice system. Recently, considerable attention has been devoted to the question of whether such tools may suffer from racial bias. In this type of assessment, a fundamental issue is that the training and evaluation of the model is based on a variable (arrest) that may represent a noisy version of an unobserved outcome of more central interest (offense). We propose a sensitivity analysis framework for assessing how assumptions on the noise across groups affect the predictive bias properties of the risk assessment model as a predictor of reoffense. Our experimental results on two real world criminal justice data sets demonstrate how even small biases in the observed labels may call into question the conclusions of an analysis based on the noisy outcome.  
\end{abstract}


\section{Introduction}
The goal of recidivism risk assessment instruments (RAI's) is to estimate the likelihood that an individual will reoffend at some future point in time, such as while on release pending trial, on probation or parole \citep{desmarais2013risk}.  Risk assessment tools have long been used in the criminal justice system to guide interventions aimed at reducing recidivism risk \citep{james2015risk}. More recently they have received considerable attention as major components of broader pretrial reform efforts seeking to reduce unnecessary pretrial detention without compromising public safety. From a public safety standpoint, society incurs a cost when a crime is committed, irrespective of whether the crime results in an arrest.  The relevant fairness question in this context is thus whether a tool provides an ``unbiased'' prediction of who goes on to commit future crimes. However, because offending is not directly observed, risk assessment models are trained and evaluated on data where the target variable is rearrest, reconviction, or reincarceration.

While these observed proxies for offending may be of interest in their own right, they are problematic as a basis for predictive bias assessment, particularly with respect to race.  Racial disparities in rearrest rates may stem from two separate causes: differential \textit{involvement} in crime, and differential law enforcement practices, also known as differential \textit{selection}  \citep{piquero2008assessing}. Rearrest is a result of not only an individual's actions, but also of law enforcement practices affecting the likelihood of getting arrested for crimes committed (or even for crimes not committed).  The limited evidence that exists suggests that differential law enforcement is not a major factor in arrests for violent crimes \citep{piquero2015understanding}. Problematically, though, for lower level offenses, which form the majority of arrests in existing data, there is reason to believe that the likelihood of getting arrested for a committed offense does differ across racial groups.  Evidence of differential selection is strongest in the case of drug crimes, where surveys suggest that whites are at least as likely as blacks to sell or use drugs; yet blacks are more than twice as likely to be arrested for drug-related offenses \citep{rothwell2014war}. This \textit{racially differential} discrepancy between the unobservable outcome $Y^*$ (reoffense) and the noisy observed variable $Y$ (rearrest) poses a critical challenge when evaluating RAI's for racial predictive bias. In this paper, we will refer to such differential discrepancy as \emph{target variable bias} (TVB).  As we show, in the presence of TVB, a model that appears to be fair with respect to rearrest could be an unfair predictor of reoffense.  

We develop a statistical sensitivity analysis framework for evaluating RAI's according to several of the most common fairness metrics, including calibration, predictive parity, and error rate balance.  Our approach is conceptually inspired by sensitivity analysis approaches widely used in causal inference studies \citep{rosenbaum2014sensitivity}.   When presenting analytic results it is common to report not only point estimates and confidence intervals, but also a parameter $\Gamma$ reflecting the magnitude of unobserved confounding that would be sufficient to nullify the observed results.  In this work we introduce a similar parameter, $\alpha$, that governs the level of label bias in the observed data.  Our methods characterize how the fairness properties of a model vary with $\alpha$, and can be used to determine the level of label noise sufficient to contradict the observed findings about those properties.  We illustrate our approach through a reanalysis of the fairness properties of the COMPAS RAI used in the ProPublica debate, and a risk assessment tool developed on data provided by the Pennsylvania Commission on Sentencing.  

\subsection{Related work}

What we call target variable bias is often referred to as \textit{differential} outcome measurement bias or differential outcome misclassification bias in the statistics and epidemiology literature on measurement error \citep{carroll2006measurement, grace2016statistical}. Most of the measurement error literature is concerned with the problem of \textit{non}-differentially mismeasured exposure (treatment), covariates, and outcomes. That is, while this form of data bias has a name, it has received little attention relative to other measurement issues.  The work of \cite{imai2010causal} is a notable exception. They do consider the setting of differential measurement error, but their goal is different from ours in that they are seeking to estimate a causal effect parameter. 

In the machine learning literature, our setting is known as \textit{censoring positive and unlabeled (PU) learning} \citep{menon2015learning}. This literature differs from the current work in two key ways.  First, while the case of feature-independent noise has been widely studied \citep{elkan2008learning, scott2009novelty, du2014analysis, liu2016classification, menon2015learning}, our work contributes to the nascent literature on feature-dependent noise \citep{menon2016learning, bekker2018learning, scott2018generalized, bootkrajang2018towards, cannings2018classification, he2018instance}.  We believe our paper is among the first to consider issues of fairness in the context of PU learning.  

There are also connections between the goal of our work and causal approaches to algorithmic bias that have recently been proposed in the fairness literature \citep{kusner2017counterfactual,loftus2018causal,kilbertus2017avoiding,nabi2018fair}. These works provide an approach to addressing biases in the observed data by attempting to directly model the causal structure governing the data generating process. Problematically, the underlying assumptions are often not empirically testable, and when violated may result in incorrect inference.  

Lastly, label noise has been briefly mentioned in prior work as a potential concern in the training and evaluation of RAI's \citep{johndrow2016fairness,Corbett-Davies:2017:ADM:3097983.3098095,corbett2018measure}.  However, none of these works undertake a formal analysis of how label noise affects training or evaluation. 


\section{Problem setup}\label{sec:setup}

We denote the observed noisy outcome (e.g., rearrest) by $Y$, the true unobserved outcome (e.g., reoffense) by $Y^*$, the set of covariates (e.g. age, criminal history) by $X$, the group indicator (race) by $A\in \{b,w\}$, and the risk score (our RAI) by $S = S(X,A)$.  The risk score $S(x,a)$ can be thought of as an empirical estimate of $\E[Y|X=x,A=a]$.  When discussing binary classification metrics, we will set a risk threshold $s_{HR}$ applied to $S$ to obtain the classifier $\hat Y = \mathds{1}_{S > s_{HR}}$. 
The discrepancy between the observed and true outcome is captured in the \textit{noise rate} function  $\gamma(x,a,y)\coloneqq \Pb(Y=1-y|X=x,A=a,Y^*=y)$.  \textbf{A central aim of this work is to characterize what can be learned about the predictive bias properties of $S$ as a predictor of the true unobserved outcome $Y^*$ under assumptions on the magnitude but not the structure of the noise.}

We make two simplifying assumptions that, while implausible in practice, greatly simplify exposition in the main manuscript and reduce the notational overhead.  First, we assume that the noise is one-sided, which rules out the case of ``false arrests.''  
\begin{assumption}\label{ass:noise_process}
    $\gamma(x,a,0)=0$ for all $x$ and $a$.
\end{assumption}
 This allows us to drop the dependency on $Y^*$ in the notation of $\gamma$, and rewrite $\E[Y | X=x, A=a]$ as $(1-\gamma(x, a))\E[Y^* |X=x, A=a]$.  That is, the discrepancy between $Y$ and $Y^*$ is due to the presence of ``hidden recidivists''. 
Table~\ref{tab:cm_hidden} describes the general setup for this setting. The left table represents the observed confusion matrix expressed in terms of the cell frequencies $p_{ij} = \P(Y = i, \hat Y = j)$; the right table introduces the parameters $\alpha_{j} \coloneqq \Pb(Y^*=1, Y=0, \hat Y=j)$. Large values of $\alpha_1$ indicate that hidden recidivists are \textit{more} likely to be classified as high risk, while large values of $\alpha_0$ indicate that hidden recidivists are \textit{less} likely to be classified as high risk. We also define $\alpha \coloneqq \alpha_0 + \alpha_1 =\E Y^* - \E Y$ that corresponds to the overall proportion of ``hidden recidivists'' in the observed data.
\begin{table}[t]
\centering
\resizebox{\columnwidth}{!}{%
\begin{tabular}{|l|cc|}
\hline
~ & $\hat Y = 0$ & $\hat Y = 1$ \\ \hline
$ Y = 0$ & $p_{00}$ & $p_{01}$ \\ 
$ Y = 1$ & $p_{10}$ & $p_{11}$ \\
\hline
\end{tabular}
\quad 
\begin{tabular}{|l|cc|}
\hline
~ & $\hat Y = 0$ & $\hat Y = 1$ \\ \hline
\\ [-1em]
$Y^* = 0$ & $p_{00} - \alpha_0$ & $p_{01} - \alpha_1$ \\ 
$Y^* = 1$ & $p_{10} + \alpha_0$ & $p_{11} + \alpha_1$ \\
\hline
\end{tabular}
}
\caption{Observed (left) and true (right) confusion matrices for arrest/offense and predicted risk.}
\label{tab:cm_hidden}
\end{table}

Second, in the main paper we suppose that one of the groups is being observed \emph{without bias}.
\begin{assumption}\label{ass:no_noise_black}
    $\gamma(x,b,1)=0$ for all $x$.
\end{assumption}
 That is, for $A = b$ we assume that $ Y^* = Y$. In the running COMPAS example, this amounts to operating as though we observed the true offenses for the black population.  One could also think of $\gamma$ as capturing the \textit{additional} degree of hidden recidivism in the white population relative to the black population. Again, this assumption is made solely to simplify exposition, and it does not qualitatively affect the presented results. \footnote{For this reason, in the paper we typically denote $\alpha\coloneqq \E[Y^*|A=w]-\E[ Y|A=w]\coloneqq\alpha^w$.}  
 In Supplement \textsection \ref{app:noise_both_groups} we show how all results are readily extensible to the case where this assumption is removed.



As we shall show next in Section \ref{sec:evaluation}, most of the bounds in our sensitivity analysis correspond to the case where the hidden recidivists correspond to the highest/lowest-scoring ($\alpha_0 \text{ or } \alpha_1=0$) defendants for whom we observed $Y = 0$.  While these extreme cases may seem unlikely in practice, they generally cannot be ruled out on the basis of the observed data alone without further assumptions.  
In such settings, existing methods typically (1) assume some data generating mechanism to conduct sensitivity analysis \citep{heckman1979sample, little2019statistical, robins2000sensitivity, molenberghs2014handbook}, (2) assume parametric models and estimate the noise by EM algorithms \citep{rubin1976inference, bekker2018learning}, or (3) impose stronger conditions on the noise processes. For instance, $\gamma$ may be assumed to depend only on a subset of $X$ \citep{bekker2018learning} or be a monotonic function of $\E[Y|X=x]$ \citep{menon2016learning, scott2018generalized}.  

\textbf{In this paper we are primarily interested in what can be said about the predictive bias properties of an RAI without untestable structural assumptions on the noise process}.  We note, however, that our results can be adapted to incorporate structural assumptions when reasonable ones are available. 
 For instance, an assumption tailored to our setting might be $ Y \ind X \mid (Y^* ,A)$.\footnote{This is a slight modification of label-dependent noise, or noise at random. In the PU learning and missing data literature, the latter is known as {\it selected at random (SAR)} \citep{bekker2018learning} and {\it missing not at random (MNAR)} \citep{rubin1976inference} respectively.}   This would assume that the noise process is constant within groups.  Such an assumption probabilistically rules out extreme cases for $\alpha_0$ and $\alpha_1$, and, as we show in Supplement \textsection\ref{sec:estimation_race_lab_noise}, it allows us to obtain tighter estimation results.
There we also demonstrate how a range of results from the label-dependent noise literature can be easily adapted to our setting.

\subsection{Data and background} \label{sec:propublica}
In May 2016 an investigative journalism team at ProPublica released a report on a proprietary risk assessment instrument called COMPAS, developed by Northpointe Inc (now Equivant)\citep{propublica2016}. The investigation found that the COMPAS instrument had significantly higher false positive rates and lower false negative rates for black defendants than for white defendants.  This evidence led the authors to conclude that COMPAS is biased against black defendants. 
The report was met with a critical response challenging its central conclusion \citep{floresfalse,dieterich2016compas,corbett-davies-2016}.  Error rate imbalance, critics argued, is not an indication of racial bias.  Instead, RAI's should be assessed for properties such as predictive parity \citep{dieterich2016compas} and calibration\citep{floresfalse}, which COMPAS was shown to satisfy.
A series of papers reflecting on the debate showed that when recidivism prevalence varies across groups, as is observed to be the case in ProPublica's Broward County data, a tool cannot simultaneously satisfy both predictive parity (calibration) and error rate balance (resp. balance for the positive and negative class) \citep{kleinberg2016inherent,chouldechova2017fairlong,berk2017fairness}.

One popular interpretation of such ``impossibility results'' is that error rate imbalance is a (perhaps inconsequential) artifact of differences in recidivism (rearrest) prevalence across groups.  That is, if one were to assess the instrument on a population where prevalence was equal, the RAI could (might be expected to) achieve parity on all of the metrics simultaneously.  Applying our framework to reanalyse the data in the setting where true offense rates are assumed to be the same across groups, we show that disparities with respect to $Y^*$ (reoffense) may in fact be \textit{greater} than those observed for $Y$ (rearrest). 


We also analyze a second private data set provided by the Pennsylvania Sentencing Commission for the purpose of research.  This dataset contains information on all offenders sentenced in the state's criminal courts between 2004-2006.  
In reports published by the Commission, they observe that the risk assessment tool they constructed appeared to overestimate risk for white offenders.  
While we do not have access to their tool, the tool we construct by applying regularized logistic regression to their data evidences the same miscalibration issues.  Our empirical results are based on applying this score to a held out set of $55031$ offenders, of whom 65.4\% are white.

\section{Sensitivity analysis under target variable bias}\label{sec:evaluation}
This section presents our main technical results, coupled with experiments that demonstrate how the results may be used in practice. All proofs are contained in Supplement \textsection \ref{appendix:sensitivity_proofs}.  Given observations $( Y, S)$ and a classification threshold $s_{HR}$, we want to understand how the relationship between the observed ($M$) and unobserved ($M^*$) performance metrics depends on the noise level $\alpha$ in the problem setup outlined in Section 2.  Superscripts $w$ and $b$ denote within-race group estimates. We present sensitivity analysis results for predictive parity, error rate balance (aka equalized odds \citep{hardt2016equality}), accuracy parity, and two tests of differential calibration.  Supplement~\textsection\ref{sec:fairpromote} presents experiments on the COMPAS data set for two fairness-promoting algorithms.  All code is available at \url{https://github.com/ricfog/Fairness-tvb}.

\subsection{Error rate balance and predictive parity} 
We begin by presenting results for the false positive rate ($FPR$), the false negative rate ($FNR$), and the positive predicted value ($PPV$).  Our first result shows that the observed values $FPR$ and $FNR$ impose constraints on the true error rates even if no assumptions are made on the magnitude of the noise.

\begin{proposition}\label{prop:agnostic_bounds}
Suppose that $1-FPR<FNR$. Then $FNR\leq FNR^*$ and $FPR\geq FPR^*$ cannot both hold. If $1-FPR > FNR$, then the opposite inequalities can not both hold.
\end{proposition}
\vspace{-0.5em}
Proposition \ref{prop:agnostic_bounds} permits us to rule out one of the possible relations between observed and true error rates based solely on observed quantities. 
\begin{mdframed}
\excompas{}  In ProPublica's COMPAS analysis, we observe that $FPR^w = 0.23$ and $FNR^w = 0.48$.  We are thus in the case where $1 - FPR > FNR$, and therefore either $FNR^w = 0.48 \le FNR^{*w}$ or $FPR^w = 0.23 \ge FPR^{*w}$, or both.   
\end{mdframed}

\begin{figure*}
\centering
  \begin{subfigure}[t]{0.48\textwidth}
  \centering
    \includegraphics[width=0.8\textwidth, height=3cm]{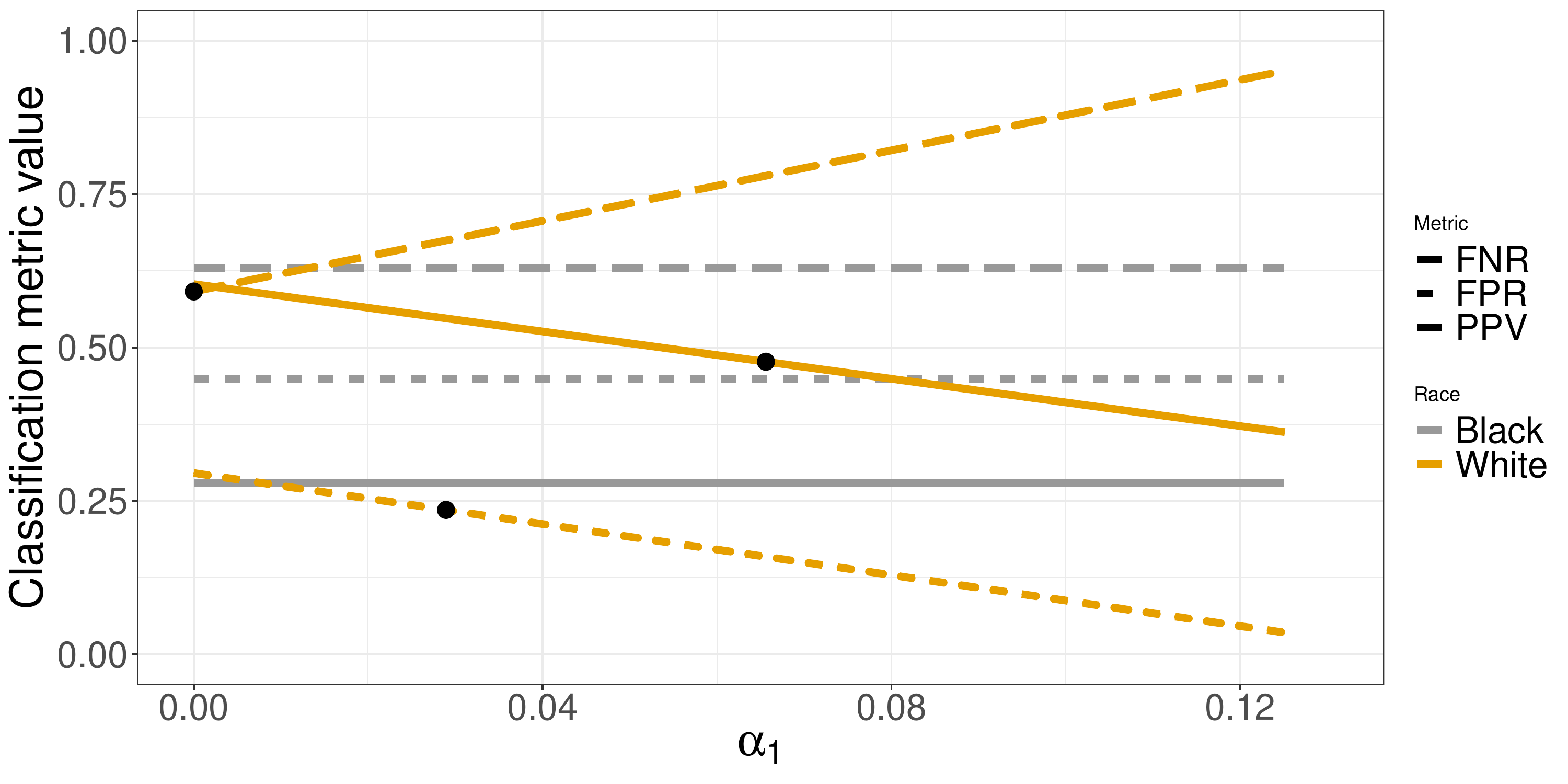}
    \caption{$\alpha$ fixed at $0.12$ to equalize reoffense rates across groups. Black dots shown indicate the value $(\alpha_1, M^*)$ for which $M^*(\alpha_1) = M$ (observed equals true).}
 \label{fig:metric-disp-plot}
  \end{subfigure}\hfill
  \centering
  \begin{subfigure}[t]{0.48\textwidth}
  \centering
   \includegraphics[width=0.8\textwidth, height=3cm]{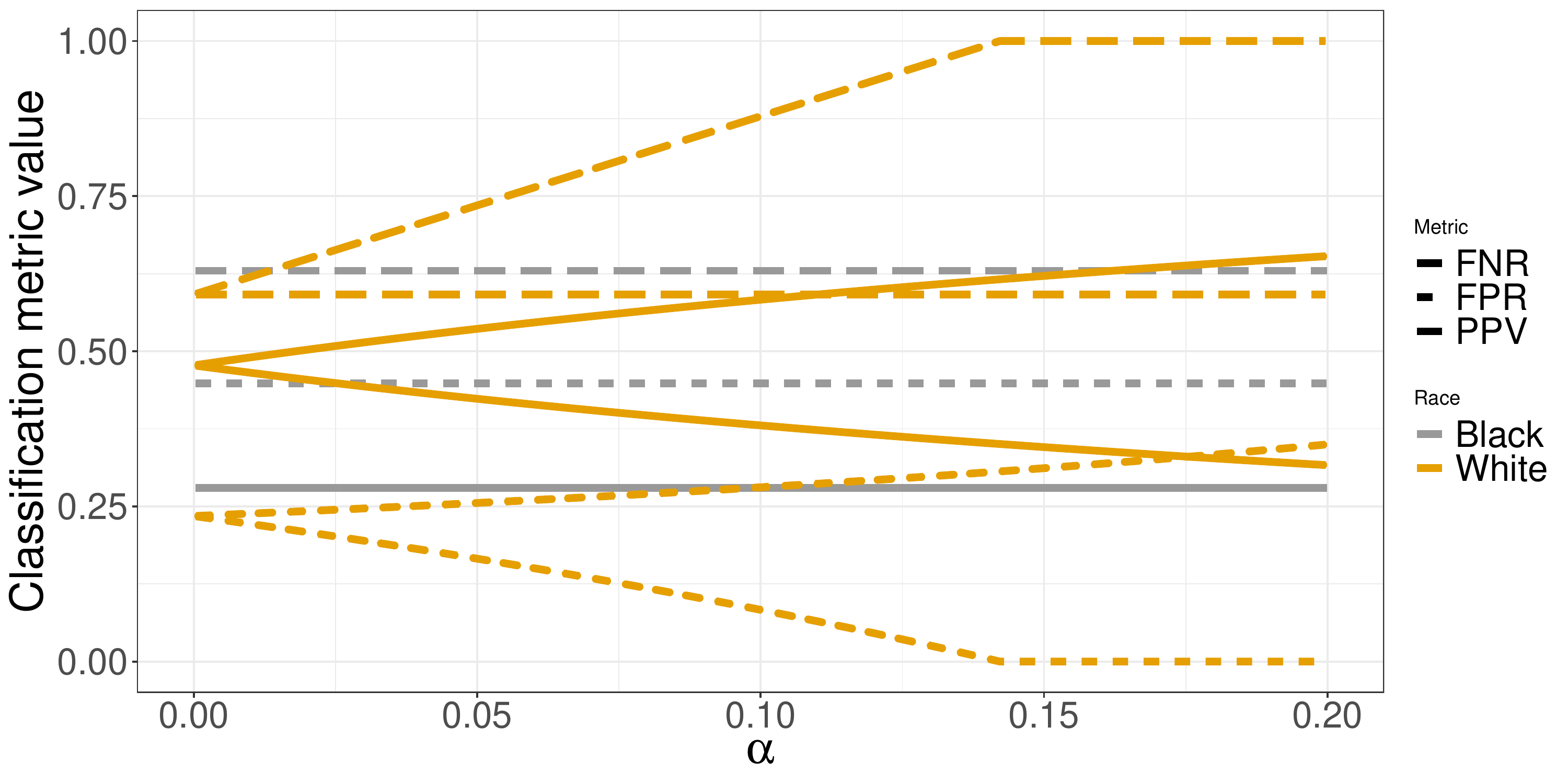}
   \caption{Bounds as described in Theorem \ref{prop:fpr_fnr_ppv_bounds} in terms of $\alpha$. Red area corresponds to region where $\text{sign}(M^{*w}-M^b)\neq \text{sign}(M^w-M^b)$.}
\label{fig:bounds_error_rates}
  \end{subfigure}
  \caption{Analysis of predictive parity and error rates for COMPAS across different TVB scenarios. Orange lines show values of $FPR^{*w}$, $FNR^{*w}$, and $PPV^{*w}$.  Grey lines show corresponding values for the black population.}
\end{figure*}

The next set of results directly relate the observed metrics $M$ to the target quantities $M^*$ based on the noise level $\alpha$. Table 1 summarizes the relationship between the observed and target confusion tables used to derive these relationships. While a version of the $FPR$ results was previously reported in \citep{claesen2015assessing}, the case of $PPV$ and $FNR$ are novel.

\begin{theorem}\label{prop:fpr_fnr_ppv_bounds}
Under the setup of Table~\ref{tab:cm_hidden}, the target values $FPR^*$, $FNR^*$, and $PPV^*$ can be sharply related to observed quantities as follows:
\begin{gather}
\frac{p_{01}-\alpha}{p_{00}+p_{01}-\alpha}\leq FPR^*(\alpha_0, \alpha_1) 
\leq \frac{p_{01}}{p_{00}+p_{01}-\alpha} \label{eq:fpr_bound}\\ 
\frac{p_{10}}{p_{10}+p_{11}+\alpha}\leq FNR^*(\alpha_0, \alpha_1) 
\leq \frac{p_{10}+\alpha}{p_{10}+p_{11}+\alpha}\label{eq:fnr_bound} \\
PPV\leq PPV^*(\alpha_0, \alpha_1) 
\leq \frac{p_{11}+\alpha}{p_{01}+p_{11}}\label{eq:ppv_bound}
\end{gather}
\end{theorem}

\begin{mdframed}
\excompas{} This result allows us to reanalyse ProPublica's COMPAS data to answer the question: {\it If the reoffense rate was equal across races, would disparities disappear?}  Figure \ref{fig:metric-disp-plot} shows the possible values of $PPV(\alpha_0, \alpha_1)$, $FPR(\alpha_0, \alpha_1)$,  and $FNR(\alpha_0, \alpha_1)$ for fixed $\alpha = 0.12$.
At this choice of $\alpha$, the true reoffense rate among white defendants is assumed equal to the rate observed for black defendants. Since $\alpha$ is fixed, $\alpha_0 = 0.12 - \alpha_1$ and hence the metrics are a function of just $\alpha_1$. We see that \emph{for most values of $\alpha_1$ disparities are even greater than what is observed}. Furthermore, while there exist values of $\alpha_1$ under which the true metric for white defendants would equal the observed (and assumed true) metric for black defendants, the equalizing value of $\alpha_1$ differs across the metrics.  

Figure \ref{fig:bounds_error_rates} shows the theoretical bounds (orange lines) provided by Theorem \ref{prop:fpr_fnr_ppv_bounds} as functions of $\alpha$ for the white population, and the observed metrics for the black population (grey lines) on the COMPAS data. We highlight the regions highlighted in red, which indicate areas where the true disparity in metrics could be of a different sign than what is observed.  This plot also shows that parity on the true $FPR$ and $FNR$ is infeasible in this data at the given choice of classification threshold.
\end{mdframed}


As a corollary of this result we can also study the question: \textit{Under what level of label noise could we expect disparities on a given metric to be smaller in truth than what was observed?}  First, note that when the observed recidivism rate is greater in group $b$ than $w$, as in the case of the COMPAS example, we will generally observe $FPR^w \le FPR^{b}$ and $FNR^w \ge FNR^{b}$.  A necessary condition for the disparity between the true error rates to be no larger than that for the observed rates is thus that $FNR^w \ge FNR^{*w}(\alpha_0, \alpha_1)$ and $FPR^w \le FPR^{*w} (\alpha_0, \alpha_1)$.  The following corollary characterizes when this occurs.  
\begin{corollary}\label{prop:fpr_fnr_unknown_bounds}
In the notation of Theorem~\ref{prop:fpr_fnr_ppv_bounds},
\begin{align}
FPR \ge \frac{\alpha_1}{\alpha} & \iff FPR \le FPR^*(\alpha, \alpha_1) \label{eq:condition_fpr}\\
FNR \ge \frac{\alpha_0}{\alpha} & \iff FNR \ge FNR^*(\alpha, \alpha_1) \label{eq:condition_fnr},
\end{align}
with equality on LHS iff there is equality on RHS.
\end{corollary}


The condition in \eqref{eq:condition_fnr} turns out to be equivalent to the odds ratio:\footnote{\citep{kallus2018residual} obtain similar expressions in their study of ``residual unfairness'' in the context of a related data bias problem. They consider the setting where we fail to observe outcomes entirely for a fraction of the population (e.g., defendants who are not released on bail, and thus do not have the opportunity to recidivate).  When viewed as functions of the underlying classification threshold $s_{HR}$, these odds ratios are interpreted in \citep{kallus2018residual} as a type of stochastic dominance condition.}
\begin{gather}\label{eq:or_fnr}
\frac{\Pb(\hat{Y}=1|Y^*=1,Y =0)/ \Pb(\hat{Y}=0|Y^*=1,Y=0)}{\Pb(\hat{Y}=1|Y=1)/\Pb(\hat{Y}=0|Y=1)}\geq 1.
\end{gather}
This condition tells us that (\ref{eq:condition_fnr}) holds precisely when the odds of correctly classifying a \textit{hidden} recidivist to $\hat Y = 1$ are greater than the odds of correctly classifying an \textit{observed} recidivist, which seems unlikely to hold in practice. A similar interpretation can be derived for $FPR$:  condition \eqref{eq:condition_fpr} holds when the odds of misclassifying a \textit{hidden} recidivist to $\hat Y = 0$ are higher than those of correctly classifying an \textit{observed} non-recidivist.

\begin{mdframed}
\excompas{} Conditions \eqref{eq:condition_fpr} and \eqref{eq:condition_fnr} in Corollary \ref{prop:fpr_fnr_ppv_bounds} require $\alpha_1\leq 0.3\alpha_0$ and $\alpha_1\geq 1.09\alpha_0$ respectively.  Note, however, that both conditions cannot simultaneously hold, as formally shown in Proposition \ref{prop:agnostic_bounds}.
\end{mdframed}

In practice, if the predicted risk for hidden recidivists was generally low, condition (\ref{eq:or_fnr}) would likely not hold. Consequently, we would thus have $FNR^w - FNR^b \le FNR^{*w}(\alpha, \alpha_1) - FNR^b$, which says that the true $FNR$ disparity between groups would be greater than the observed $FNR$ disparity. 

\begin{figure*}[t]
\centering
  \begin{subfigure}[t]{0.45\textwidth}
   \includegraphics[width=\linewidth]{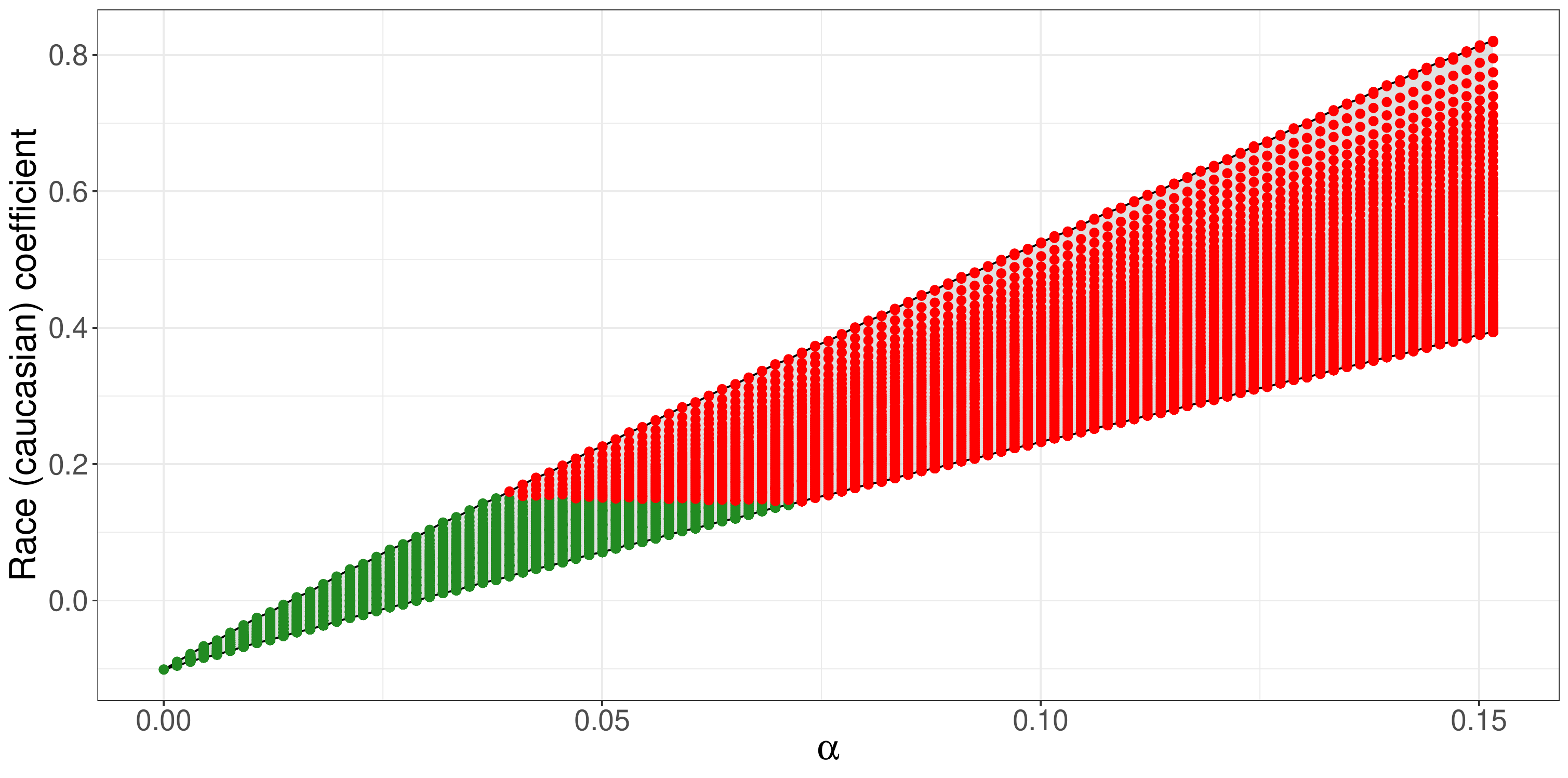}
   \caption{Calibration analysis on COMPAS data.}
  \end{subfigure} \hspace{2em}
  \begin{subfigure}[t]{0.45\textwidth}
   \includegraphics[width=\linewidth]{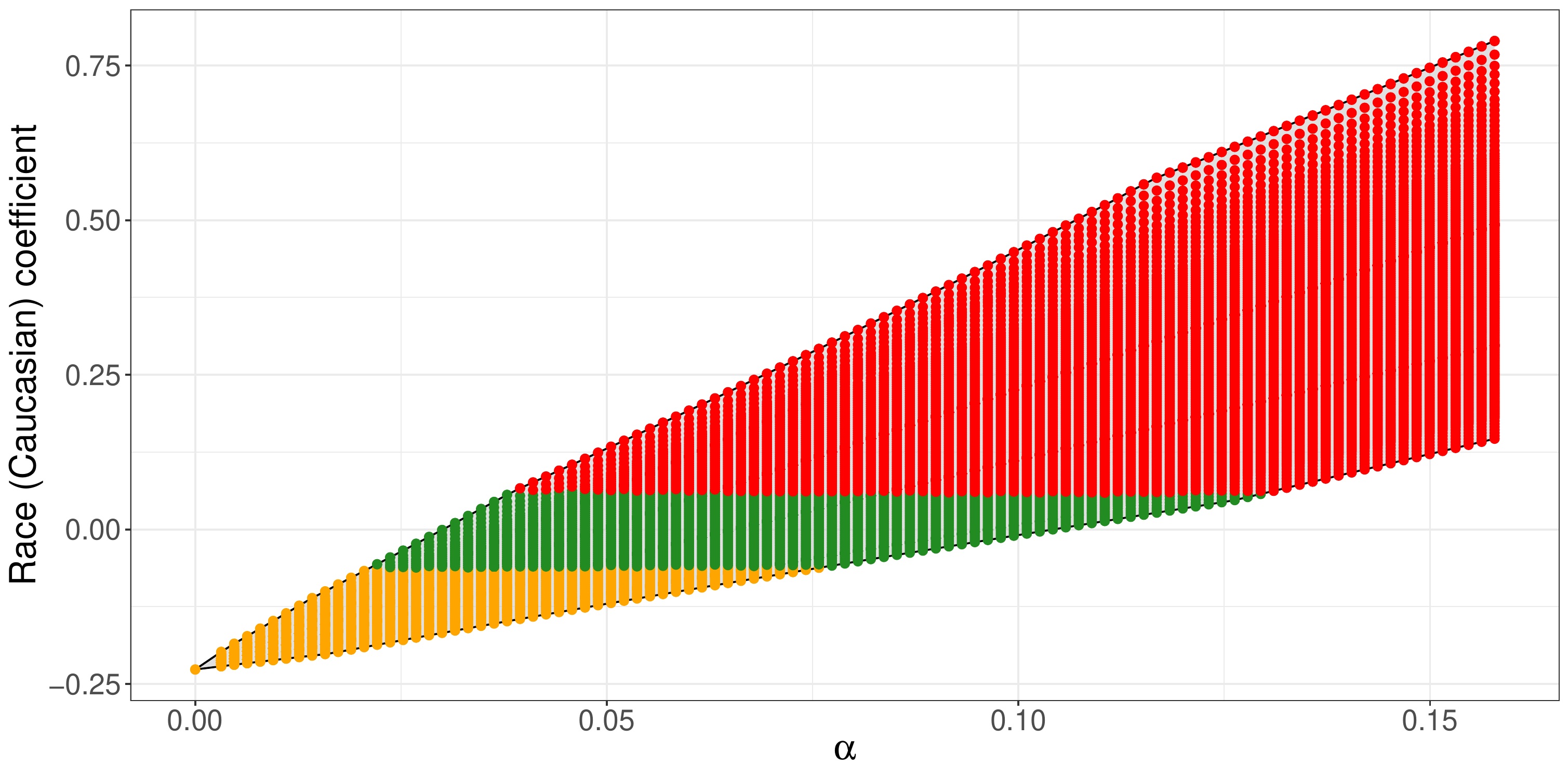}
 \caption{Calibration analysis on Sentencing comm. data.}
  \end{subfigure} \hspace{2em}
\caption{Sensitivity analysis for the race coefficient in a logistic regression test of calibration as described in Section~\ref{sec:calib_logistic}. Green region indicates the race coefficient is not statistically significant for testing calibration wrt \textit{offense}. Red (resp., orange) region indicates statistically significant bias against the black (resp., white) group. }\label{fig:bounds_coefficient}
\end{figure*}

\subsection{Accuracy equity} 
In their response to the ProPublica investigation, \cite{dieterich2016compas} demonstrated that COMPAS satisfies predictive parity (equality of $PPV$ and $NPV$ across groups), and what they term \textit{accuracy equity} (equality of $AUC$).  \cite{menon2015learning} and \cite{jain2017recovering} previously considered estimation of the AUC under label noise, but in the simpler setting of label-dependent noise. Here we obtain bounds for the true AUC in the general instance-dependent noise setting through its relation to the Mann-Whitney U-statistic. 

Let $n_{y}= \#\{Y_i = y\}$ denote the number of observations with outcome $Y = y \in \{0,1\}$.  We will assume that there are $k = \ceil{n \alpha}$ hidden recidivists present in the observed data, with $k < \min(n_0, n_1)$.    Let $r_i$ denote the adjusted\footnote{In the case of ties among the scores, the U-statistic is calculated using fractional ranks.} rank of observation $i$ when ordered in ascending order of the score $S$.  Lastly, let $R_1 = \sum_{i : Y_i = 1}r_{i}$ denote the sum of the ranks for observations in class $Y = 1$.  In this notation, the observed $AUC$ of $S$ is given by
\begin{equation}
AUC = \frac{R_1}{n_1(n-n_1)} - \frac{n_1 + 1}{2(n - n_1)}
\end{equation}
Let $L_{0,k}$ denote the indexes of the lowest-ranked (i.e., lowest-scoring) observations in class $Y = 0$.  Likewise, let $H_{0,k}$ denote the indexes of the highest-ranked (i.e., highest-scoring) observations in class $Y = 0$.  

\begin{proposition}\label{prop:bounds_auc}
In the presence of $k$ hidden recidivists, the target value AUC is bounded as follows:
\begin{align}\label{eq:bounds_AUC}
\frac{R_{1}+\sum_{i \in L_{0,k}}r_i -\beta_k}{(n_{0}-k)(n_{1}+k)} \leq AUC^* &\le \frac{ R_{1}+\sum_{i \in H_{0,k}}r_i-\beta_k}{(n_{0}-k)(n_{1}+k)}
\end{align}
where $\beta_{k}=(n_{1}+k)(n_{1}+k +1)/2$. 
\end{proposition}
It is easy to see that the upper and lower bounds correspond to the settings where the hidden recidivists are, respectively, the highest and lowest scoring defendants with $Y = 0$. This result tells us, for instance, that if the hidden recidivists are more likely to have high scores, then the true $AUC$ will be greater than the observed $AUC$.  One key difference between the AUC result and the previous analysis of error metrics is that now the impact of label noise depends on the ranks of the hidden recidivists, and not only on the dichotomized version of the risk score.

\begin{mdframed}
\excompas{} The observed AUC for both the black and white defendant population is around $0.69$.  Evaluating the bounds from the proposition for the white population, we find that for $\alpha=0.05$ and $\alpha=0.12$, the $AUC^{*w}$ is bounded between $[0.63, 0.76]$ and $[0.51, 0.84]$, respectively.  These bounds are very wide, but they can be narrowed if we are willing to make further assumptions on the likely ranks of the hidden recidivists.  
\end{mdframed}

\subsection{Calibration testing via logistic regression} \label{sec:calib_logistic}
One of the most common metrics for assessing predictive bias of RAI's is a test of \textit{calibration} or \textit{differential prediction} \citep{skeem2015risk}. Formally, we say that a risk score $S$ is well-calibrated with respect to $A$ if 
\begin{equation}
\E[Y \mid S=s, A = w] = \E[Y\mid S=s, A = b].  \label{eq:calib}
\end{equation}
for all values of $S$. This is equivalent to requiring that $Y \ind A \mid S$. Typically calibration is assessed by running a logistic regression and testing for statistical significance of $A$ in $ Y \sim S$ vs. $Y \sim S + A$ or $Y \sim S + A + SA$ using a Wald or likelihood ratio test.\footnote{We adopt the shorthand \mbox{$Y \sim X_1 + X_2 + \cdots X_p$} to refer to the logistic regression model \mbox{$\log(p(X)/(1 - p(X)) = \beta_0 + \beta_1 X_1 + \beta_2 X_2 + \cdots + \beta_p X_p$}, where $p(X) = \P(Y = 1 \mid X)$.}  Other covariates are occasionally also included in the regression.   When the coefficients of $A$ are not statistically significant, $S$ is deemed to be well-calibrated with respect to $A$. This approach was taken  by \cite{floresfalse} to confirm racial calibration for the COMPAS RAI. Note that in the presence of TVB, such tests provide evidence that $S$ is well-calibrated as a predictor of $Y$ (rearrest).  We wish to understand what this means about $S$ as a predictor of the true outcome $Y^*$ (reoffense).  Our main result is as follows.    

\begin{figure*}[t]
\begin{subfigure}[t]{.24\linewidth}
  \centering
\includegraphics[width=\linewidth]{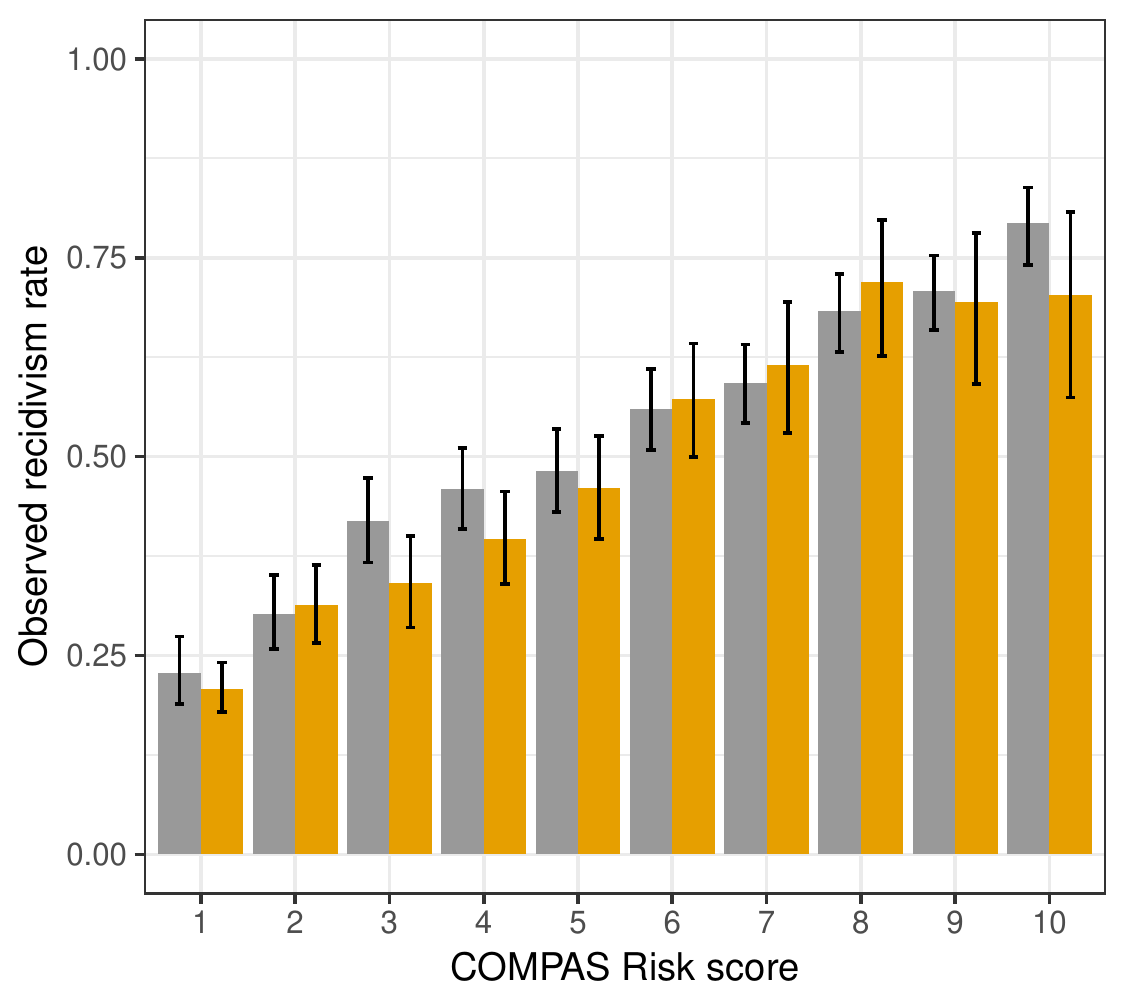}
  \caption{Observed COMPAS. \\ $T = 9.36$, $p$-value$= 0.49$.}
\end{subfigure}%
\begin{subfigure}[t]{.24\linewidth}
  \centering
\includegraphics[width=\linewidth]{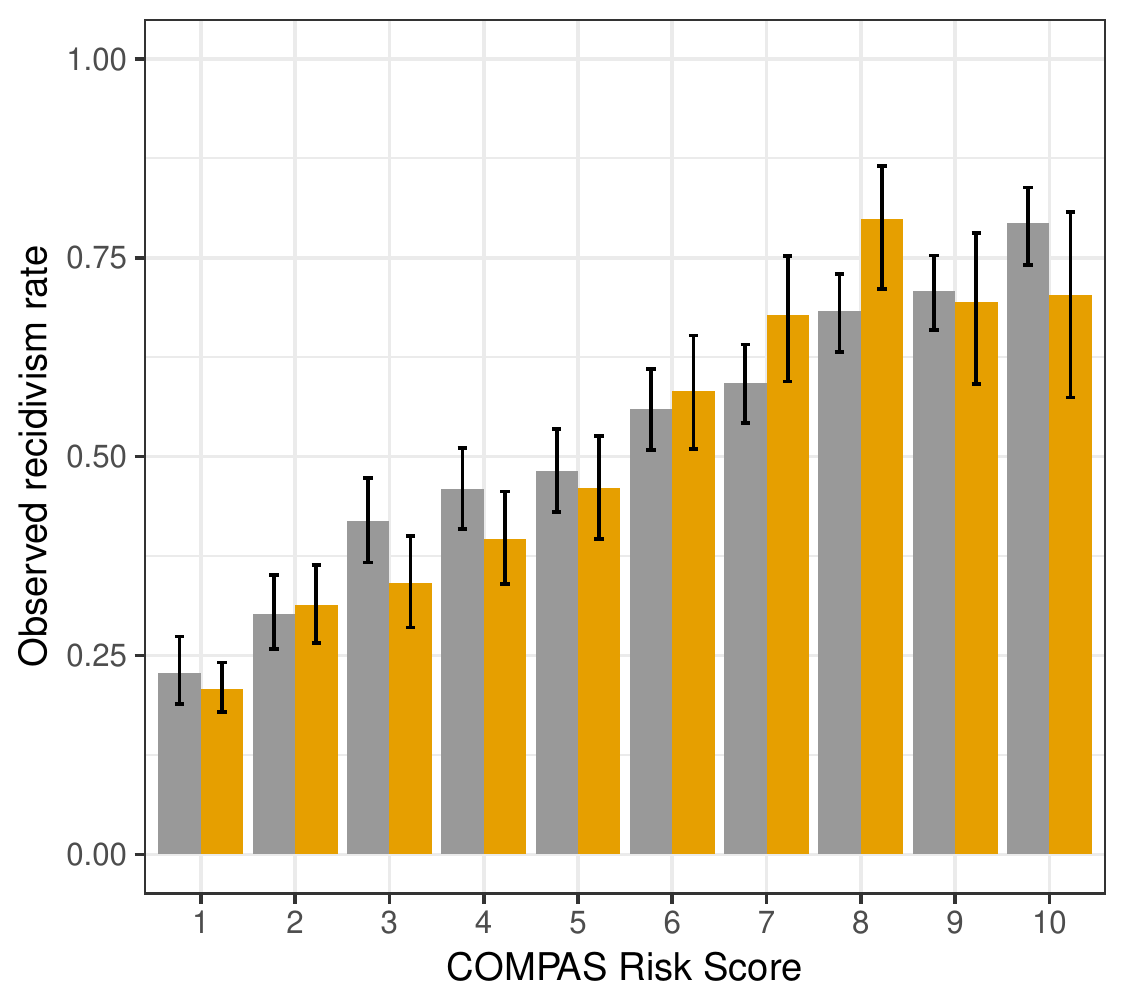}
  \caption{Minimal shift to break calibration under proportionality constraint. $N_h = 30$, $T = 19.4$, $p$-value $=0.035$.}
\end{subfigure}
\begin{subfigure}[t]{.24\linewidth}
  \centering
\includegraphics[width=\linewidth]{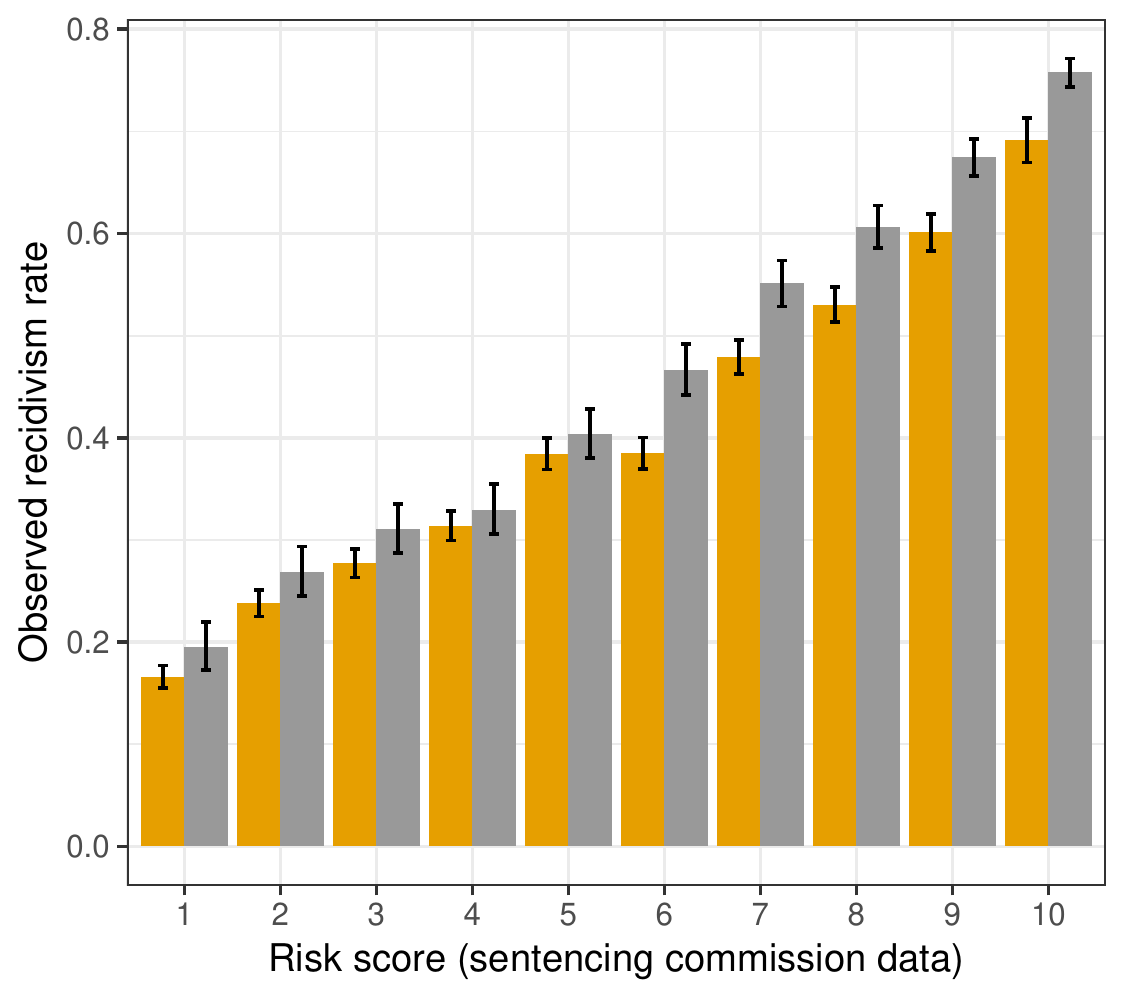}
  \caption{Observed SC score. \\ . $T = 164$, $p$-value $\approx 0$}
\end{subfigure}
\begin{subfigure}[t]{.24\linewidth}
  \centering
\includegraphics[width=\linewidth]{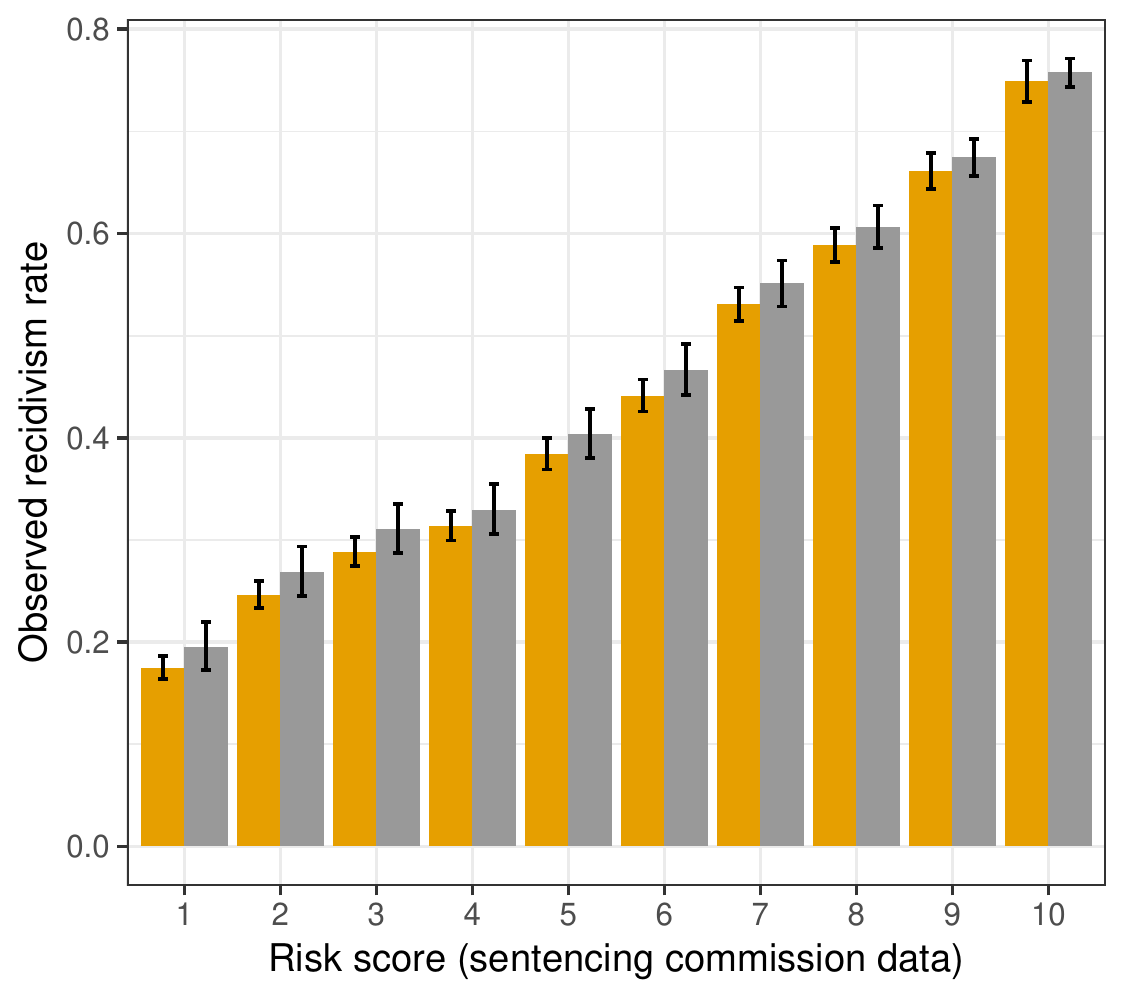}
  \caption{Minimal shift to achieve calibration, unconstrained. \\  $N_h = 1001$, $T = 18.2$, $p$-value $=0.051$}
\end{subfigure}
\caption{Sensitivity analysis for the race coefficient in a chi-squared nonparametric test of calibration as described in Section \ref{sec:chisq}.  Orange bars show white defendant data; grey bars are black defendant data.  Error bars show 95\% confidence intervals.  Plots (a) and (b) correspond to the COMPAS data example where a small amount of TVB is sufficient to lead to miscalibration.  Plots (c) and (d) are the sentencing commission (SC) example where a small amount of TVB can account for observed miscalibration in predicting arrest.  }
\label{fig:chisq_shifts}
\end{figure*}

\begin{proposition}\label{prop:logistic_bound}
Under a mild technical assumption on the design matrix,\footnote{The explanation of the assumption is deferred to the proof in the Supplement. While the assumption needs to be empirically verified case by case, in the COMPAS dataset it holds at every level of $\alpha^w$ that we considered.} for a logistic regression model of the form $Y \sim S + A$, for fixed $\alpha$, the bounds for the coefficients of $S$ and $A$ are achieved when the $\ceil{n_w \alpha}$ white defendants with the highest and lowest values of $S$ are hidden recidivists.
\end{proposition}
This result allows us to answer the question: \textit{What level of label noise $\alpha$ is sufficient to contradict the observed findings that an RAI is (or is not) well-calibrated across groups?}  We provide two illustrative examples, one where the RAI is observed to be well-calibrated as a predictor of arrest, and the other where it is not.

\begin{mdframed}
\excompas{} Figure \ref{fig:bounds_coefficient} (a) shows the feasible values for the coefficient of $A = w$ in the COMPAS data for $0 \le \alpha\le 0.16$.  The green and red areas correspond, respectively, to regions where the race coefficient \textit{is not} and \textit{is} statistically significant.  Recall that non-significance of the race coefficient indicates that the model is well-calibrated. In this analysis, we find that a TVB level as low as $\alpha = 0.07$ might be sufficient for COMPAS to fail the calibration test across \textit{all} possible noise realizations of that magnitude.  At a noise level of only $\alpha=0.04$, calibration might also fail for \textit{some} noise realizations of this magnitude. Note that our analytic results present bounds not just on the race coefficient but also on the score coefficient in the model. We present the two-dimensional bounds for the COMPAS tool in Supplement \textsection \ref{sec:extra_exp_logistic}.

\textbf{Example: Sentencing commission.} Figure \ref{fig:bounds_coefficient} (b) shows the results of the same experiment on sentencing commission (SC) data described in Section 2.1.  In absence of TVB, Figure \ref{fig:bounds_coefficient} (b) shows that this tool, unlike the COMPAS RAI, is \textit{not} observed to be well-calibrated across groups. Indeed, the coefficient for $A=w$ is statistically significantly negative, indicating the RAI overestimates risk for white offenders. Our analysis showns that TVB as low as $\alpha=0.03$ is sufficient to admit calibration. More generally, we see that for $0.03\leq \alpha\leq 0.13$ calibration might be possible for some realizations of the noise process. For a larger magnitude of TVB, the coefficient might be significant and positive; in other words, it would be possible for the instrument to underestimate the reoffense risk for the white population.
\end{mdframed}

\subsection{Calibration testing via chi-squared test}\label{sec:chisq}
We also consider the general test of conditional independence $Y \ind A \mid S$ in the setting where $S$ is either assumed to be discrete, or has been binned for the purpose of analysis. When $S$ is categorical, testing the saturated logistic model $Y \sim S + A + S A$ vs. $Y \sim S$ is precisely testing the conditional independence of $Y \ind A \mid S$. This section thus extends the analysis from the previous section beyond the (likely misspecified) simple shift-alternative considered therein. There are several asymptotically equivalent tests that can be applied to test this hypothesis \citep{hinkley1979theoretical}. We use the Pearson chi-squared test, as it is the most straightforward to analyse. 

The general setup for assessing the sensitivity of the chi-squared conditional independence test to TVB is described by Table~\ref{tab:chisq_true}.
Our goal is to understand the behavior of the chi-squared test statistic, \begin{gather}\label{eq:chisqtestt}T(h) = \sum_{k = 1}^{|S|} \sum_{a,y} \left(O_{ay}^{(k)} - E_{ay}^{(k)}\right)^2/E_{ay}^{(k)}\end{gather}
as a function of the hidden recidivist counts $h = (h_1, \ldots, h_{|S|})$.  The notations $O$ and $E$ denote the ``observed'' and ``expected'' cell counts for calculating the chi-squared statistic.  Expected counts are estimated from the data assuming the null hypothesis $Y^* \ind A \mid S$ is true.  These quantities evaluate to
\begin{align*}
O_{ay}^{(k)} &= \nk_{ay} + h_k\Ind_{a=w}(2y-1), \text{ and } \\
E_{ay}^{(k)} &= \left(\nk_{wy}+\nk_{by}+ (2y-1)h_k\right)\left(\nk_{a0}+\nk_{a1}\right)/\nk.
\end{align*}
The key observation is that, when viewed as a function of $h_k$, the numerator terms $(O_{ay}^{(k)} - E_{ay}^{(k)})^2$ are convex quadratics in $h_k$, and the denominator terms $E_{ay}^{(k)}$ are linear functions in $h_k$, constrained to be positive.  

We address two basic questions: (1) \textit{When $S$ appears racially well-calibrated for the observed $Y$, how large would $N_h$, the number of hidden recidivists, have to be for $S$ to fail the calibration test for $Y^*$?} (2) \textit{When $S$ appears to underestimate risk for the one racial group, how large would $N_h$ have to be for $S$ to appear racially well-calibrated for $Y^*$?}  Answering (1) entails maximizing the test statistic $T$ over $h$ subject to $\sum h_k \le N_h$.  Answering (2) entails minimizing the test statistic.  Note that each inner summand of equation~\eqref{eq:chisqtestt} is a quadratic-over-linear function, which is strongly convex \citep{boyd2004convex}.  The test statistic $T$ as a function of $h$ thus has the form
$T(h) = \sum_{k = 1}^{|S|} f_k(h_k)$,
where each $f_k$ is a strongly convex function. 
Since $T(h)$ is a strongly convex separable function of the $h_k$'s, the minimization can be performed with a numerical convex solver.  Note that it is also straightforward to incorporate convex constraints into the optimization.  The maximization task is a case of a separable nonlinear optimization problem, for which general tools exist.  For our analysis we instead present a practical greedy algorithm in Supplement \textsection \ref{appendix:chisq}.

\begin{table}[t]
\centering 
\begin{tabular}{|l|cc|}
\hline
$S = k$ & $Y^* = 0$ &  $Y^* = 1$ \\ \hline
$A = w$ & $n^{(k)}_{w0} - h_k$ & $n^{(k)}_{w1} + h_k$ \\ 
$A = b$ & $n^{(k)}_{b0}$ & $n^{(k)}_{b1}$ \\
\hline
\end{tabular}
\caption{Contingency table for rearrest outcome in score level $S = k \in \{1, \ldots, |S| \}$ for testing  $H_0: Y^* \ind A \mid S$ with the chi-squared test. Here $h_k$ denotes the number of ``hidden recidivists'' in the white defendant population in score level $S = k$.}
\label{tab:chisq_true}
\end{table}

\begin{mdframed}
\excompas{} Figure \ref{fig:chisq_shifts}(a) shows the observed recidivism rates for black and white defendants across the range of the COMPAS decile score.  When we apply the chi-squared test to test for calibration, we find that the COMPAS instrument appears well-calibrated with respect to race $(T = 9.36, p\textrm{-value} = 0.49)$. However, applying our method to \textit{maximize} the test statistic, we find that the presence of just $N_h = 20$ hidden recidivists is sufficient to break calibration.  This is achieved when all $N_h = 20$ hidden recidivists are located in score level 8.  Looking at the data, this is unsurprising.  Score level $8$ already has the largest observed discrepancy with the black defendant recidivism rate.  Pushing this discrepancy further will rapidly cause the test to reject. Figure \ref{fig:chisq_shifts}(b) shows the minimal shift necessary to break calibration when we impose a \textit{proportionality constraint} that prohibits allocations that concentrate too much on a single bin.  Specifically, we require that $h_k \le \epsilon n^{(k)}_{w1}$.  This ensures that the proportion of true recidivists that are hidden \textit{in any score bin} is no greater than $\epsilon$.  For our experiment we take $\epsilon = 0.1$.  Under this constraint, we find that $N_h = 30$ are sufficient to break calibration.  These are allocated as $h = (0,  0,  0,  0,  0,  12,  9,  9,  0,  0)$.

\textbf{Example: Sentencing commission.} The right panel of Figure \ref{fig:chisq_shifts} shows the observed recidivism rates for black and white defendants across the range of the decile score we constructed based on the sentencing commission data.  Unlike in the COMPAS example, we find that the SC score shows clear evidence of poor calibration $(T = 164, p\textrm{-value} \approx 0)$.  The RAI underestimates risk of rearrest for white offenders relative to black offenders across the range of score levels.  This effect is especially pronounced in the highest scores.  Applying our method to \textit{minimize} the test statistic, we find that just $N_h = 1001$ hidden recidivists are sufficient to achieve calibration.  While this may seem like a large number, there are $n_w = 31607$ white offenders in the data, of which $n_{w1} = 13552$ are observed to reoffend.  Thus the minimizing allocation requires only that $1001 / (13552 + 1001) = 6.9\%$ of all true recidivists go unobserved.  The minimizing allocation, represented in the left panel of Figure~\ref{fig:chisq_shifts}, is $h = (41,  35,  46,   0,   0, 224, 186, 197, 170, 102)$.
\end{mdframed}






\section{Conclusion}\label{sec:discussion}

When target variable bias is a concern, the sensitivity analysis framework presented in this paper can be used to quantify the level of bias sufficient to call into question conclusions about the fairness of a model obtained from biased observed data.  In the sentencing commission example, for instance, we find that a small gap in the likelihood of arrest could fully account for the observed miscalibration.  Such observations may help inform deliberations of whether to correct for observed predictive bias when doing so would further increase outcome disparities.  Furthermore, as our reanalysis of the ProPublica COMPAS data shows,  the racial disparity story goes deeper than an imbalance in observed recidivism rates.  Even if offense rates are equal across groups, the disparities could be worse with respect to offense than what is observed for arrest.  

The sensitivity analysis approach outlined in this work has generally avoided making assumptions about how the likelihood of getting caught might depend on observable features, at a cost of producing fairly wide bounds.  Existing work on self-report studies, wrongful arrests, and wrongful convictions may provide some insight into reasonable structural assumptions that may be incorporated to further refine the analysis \citep{huizinga1986reassessing,hindelang1979correlates, gilman2014understanding}. 


\bibliographystyle{plainnat}
\bibliography{mybib.bib}

\newpage
\onecolumn
\appendix
\newpage
\section*{Organization of the Supplement}
\begin{itemize}
    \item section \ref{sec:extension_2} (\textsection \ref{sec:setup}):
    \begin{itemize}
        \item motivating examples;
        \item estimators for the noise under conditions stronger than assumption \ref{ass:noise_process}.
    \end{itemize}
    \item section \ref{sec:extension_sec_3} (\textsection \ref{sec:evaluation}):
    \begin{itemize}
        \item omitted proofs for section \ref{sec:evaluation};
        \item extension of results under conditions stronger than assumption \ref{ass:noise_process};
        \item extension of results under relaxation of assumption \ref{ass:no_noise_black};
        \item further experiments.
    \end{itemize}
    \item Section C:
    \begin{itemize}
        \item experiments on error rate balance with fairness-promoting algorithms.
    \end{itemize}
\end{itemize}

\section{Extension for section 2}\label{sec:extension_2}

In this section, we use $ m = m(x)$ and $m^* = m^*(x)$ to indicate $\E[Y |X=x]$ and $\E [ Y^*|X=x]$ respectively. We also drop the dependency of $\gamma $ on $A$ and, if assumption~\ref{ass:noise_process} is used, on $Y^*$.

\subsection{Who are the likely hidden recidivists?}
In section \textsection\ref{sec:evaluation} we have argued that the worst case bounds in our sensitivity analysis occur when the hidden recidivists are either all in the low-risk bin ($\alpha_1 = 0$) or all in the high-risk bin ($\alpha_0 = 0$). Here we present two thought examples reflecting on assumption~\ref{ass:noise_process}. We show that, generally speaking, one can not rule out the ``extreme'' settings. Indeed, under assumption~\ref{ass:noise_process}, the case $m\equiv 1$ is still possible. 

$\cdot$ Example 1 $\cdot$ Suppose for instance that $X \in \{0,1\}$ is a single binary covariate, $m^*(1) = 1, m^*(0) < 0.5$, and $\gamma(1) = 0.4, \gamma(0) = 0$.  This gives $ m (1) = 0.6$ and $ m(0) = m^*_0 < 0.5 <  0.6 = m(1)$.  If we set the classification threshold at $s_{HR} = 0.5$, we would classify everyone with $X = 1$ as high-risk and everyone with $X = 0$ as low-risk.  By construction, we have $\gamma(0) = 0$, meaning that all recidivists with $X = 0$ are observed, whereas some fraction of recidivists with $X = 1$ are hidden. This in turn means that all hidden recidivists are classified as high-risk ($\alpha_0 = 0$).  A similar construction can be used to produce a case where $\alpha_1 = 0$, which corresponds to all hidden recidivists being classified as low-risk.  

\begin{figure}[t]
\centering
\includegraphics[width=0.5\linewidth]{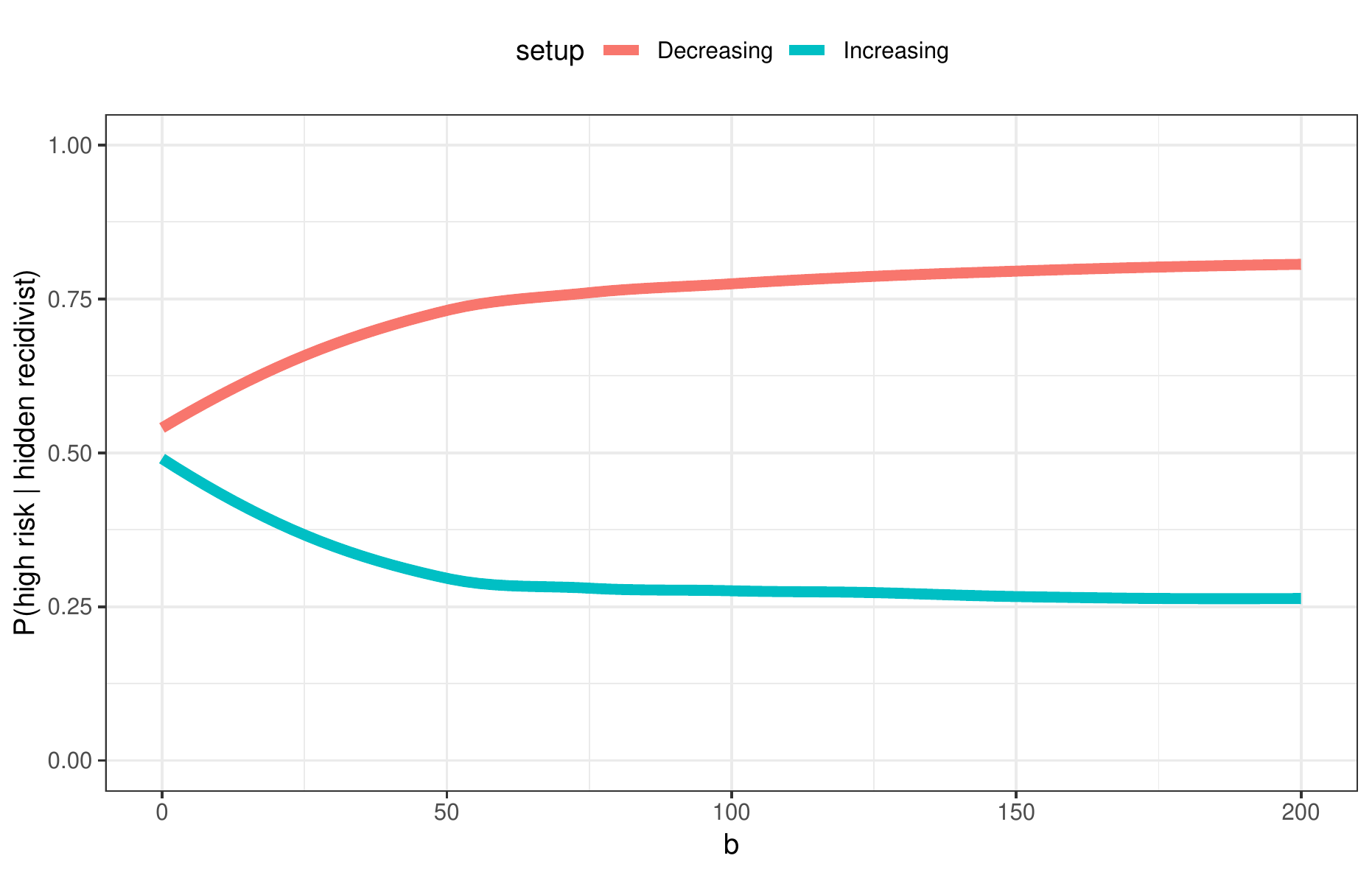}
\caption{Proportion of hidden recidivists classified as high risk under different choices of $\gamma$.  }
\label{fig:prophr_plot}
\end{figure}

$\cdot$ Example 2 $\cdot$ The first example is admittedly highly contrived and unlikely to reflect any real world scenario.  To model a more plausible scenario, we consider a setup in which we have a single feature $X \sim Unif[0,1]$, $m(x) = x$, and two forms for the likelihood of getting caught function:
\begin{gather*}
    \gamma^{Inc, b}(x) = 1-(b+1)x/(1 + bx),\\
    \gamma^{Dec, b}(x) = 1-(b+1)(1-x)/(1 + b(1-x)).
\end{gather*}
The ``Increasing'' setting $\gamma^{Inc,b}$ is one where the likelihood of getting caught increases with the likelihood of reoffense $m^*(x)$, with the functional form of the relationship governed by the parameter $b$.  The ``Decreasing'' setting has the likelihood of getting caught decreasing with the likelihood of reoffense.  We equalize the proportion of high-risk and low-risk cases by thresholding $m(x)$ at its median value in each simulation.  Figure \ref{fig:prophr_plot} shows a plot of how the fraction of hidden recidivists that get classified as high-risk varies with $b$.  Values larger than $0.5$ on this plot can be interpreted as settings where $\alpha_1 > \alpha_0$; a value of $1$, though never achieved, would correspond to the case $\alpha_0 = 0$.  This suggests that, in general, the hidden recidivists are likely to be scattered across the range of the score $S$, and are thus unlikely to concentrate entirely in the extremes of $S$.  In other words, the worst-case bounds presented in Section \ref{sec:evaluation} are, unsurprisingly, likely to be overly conservative.

\subsection{Estimation of noise}

In \textsection \ref{sec:setup}) we have argued that the assumption of constant noise is unrealistic in our setting. Indeed, in the introduction we cite the case of drug crimes (low-level offenses), where there appears to be an inconsistency in the number of arrests and users between the black and white populations; this fact might be attributed to {\it differential policing}. For other types of crimes we can imagine the effect of policing to be more similar across races.
Although we suggest to account for more complex forms of the noise, one may wish to perform a sensitivity analysis under stronger assumptions on the noise process, e.g. assume the noise to be independent of the features conditionally on the observed labels. The case of constant noise has been intensively studied during the past two decades and it is fairly well understood. In this subsection we present a simple extension of this framework to account for noise constant within groups.

\subsubsection{Estimation of one-sided label-dependent noise.}\label{sec:estimators_nar_noise}

In the paper we work under the setup of assumption \ref{ass:noise_process}, that is of one-sided feature-dependent noise.
Now, consider the following assumption.
\begin{assumption}\label{ass:label_dep_noise}
    $Y\ind X|Y^*$.
\end{assumption}
Under assumptions \ref{ass:noise_process} and \ref{ass:label_dep_noise} we refer to the noise as {\it one-sided label-dependent}. Since the noise rate $\gamma(x,1)$ is now constant, we drop the dependency on $x$ and rewrite $\gamma = \gamma(x,1)$.

We briefly describe three of the estimators for the noise rates commonly used in the literature. These estimators can be used for estimation of the noise rate in the setting of assumptions \ref{ass:noise_process} and \ref{ass:label_dep_noise}.\\

$\cdot$ Estimator 1 $\cdot$ The estimator proposed by~\citep{elkan2008learning} relies on the following assumption.
\begin{assumption}\label{ass:non_overlapping_support}(strong separability)
    $m^*(x)\in\{0,1\}$.
\end{assumption}
Then we have the following proposition.
\begin{proposition}\label{prop:strong_sep_est}
    Under assumptions \ref{ass:noise_process}, \ref{ass:label_dep_noise}, and \ref{ass:non_overlapping_support}, the following equality holds. For every $y=1$, \begin{equation} \gamma=1-m(x). \end{equation}
\end{proposition}
\vspace*{-0.2cm}
{\it Proof of proposition \ref{prop:strong_sep_est}.} 
Thanks to assumption~\ref{ass:non_overlapping_support}, $y=1$ implies $m^*(x)=1$. Consequently we have $m(x)=(1-\gamma)m^*(x)=1-\gamma$ for every $y=1$. \hfill \qed \\

Estimators 2 and 3 rely on the following assumption.
\begin{assumption}\label{ass:weak_separability}(weak separability)
    $\sup_{x}m^*(x)=1$.
\end{assumption}

$\cdot$ Estimator 2 $\cdot$ The following is also described in \citep{elkan2008learning, liu2016classification, menon2015learning}.
\begin{proposition}\label{prop:noise_nar_gamma}
    Under assumptions \ref{ass:noise_process}, \ref{ass:label_dep_noise}, and \ref{ass:weak_separability}, the following equality holds.
    \begin{equation}
        \gamma=1-\sup_x m(x)
    \end{equation}
\end{proposition}
\vspace*{-0.2cm}
{\it Proof of proposition \ref{prop:noise_nar_gamma}.}
Recall the decomposition $m(x)=(1-\gamma)m^*(x)$. Then, thanks to assumption \ref{ass:weak_separability}, we have
\begin{gather*}
    \sup_x m(x)=(1-\gamma)\sup_x m^*(x)=1-\gamma \implies \gamma=1-\sup_x m(x).
\end{gather*}  \hfill \qed\\
Consequently the rate of convergence for the estimation of $\gamma$ coincides with the one for $m(x)$.\\

$\cdot$ Estimator 3 $\cdot$ We define $\rho$, the {\it inverse noise rate}, as
\begin{gather}
\rho:=\E[Y^*|Y=0]=\frac{\alpha}{1-\E[Y]}=\frac{\gamma}{1-\E[Y]}\E[Y^*]=\frac{\gamma}{1-\E[ Y]}\frac{\E[Y]}{1-\gamma}=\frac{\gamma/(1-\gamma)}{(1-\E[Y])/\E[Y]}.
\end{gather}
Note that $\gamma$ identifies $\rho$, and vice versa. An estimator for $\rho$ has been proposed by \citep{scott2009novelty, scott2013classification}. \\
Let $q^*_{y}$ and $q_{y}$ denote the densities of $X$ conditional on $Y^*=y$ and $Y=y$ respectively. Under assumptions \ref{ass:noise_process}, \ref{ass:label_dep_noise}, and \ref{ass:weak_separability},
\begin{gather}
    \rho=\frac{\nu(q_0,q^*_1)(1-\nu(q^*_1,q_0))}{1-\nu(q^*_1,q_0)\nu(q_0,q^*_1)}
\end{gather}
where $\nu(q^*_1, q_0)=\inf_x q^*_1(x)/q_0(x)$ and $\nu(q_0, q^*_1)=\inf_x q_0(x)/q^*_1(x)$. $\nu$ corresponds to the left-derivative of the optimal ROC curve \citep{scott2009novelty}. The optimal ROC curve  is given by any scorer that is a strictly monotone transformation of $p$ \citep{clemenccon2008ranking}. In \citep{scott2009novelty} the estimator is recovered behind an assumption slightly weaker than assumption \ref{ass:weak_separability} that the authors call {\it irreducibility}; however, under this assumption, the convergence rate of the estimator is shown to be arbitrarily slow. \citep{scott2015rate} introduces an assumption equivalent to \ref{ass:weak_separability} that guarantees faster convergence rates.

It is clear that if assumption \ref{ass:weak_separability} does not hold, then the estimated noise rate is only upper bounded by $1-\sup_x m(x)$, and consequently $ m(x)\leq m^*(x)\leq m(x)+\gamma$.

\subsubsection{Estimation of one-sided race- and one-sided label-dependent noise.}\label{sec:estimation_race_lab_noise}

In our setting it is more reasonable to consider a noise process that depends on the race membership; indeed, the original motivation of our work was a concern regarding {\it differential policing} across races. To simplify notation, let $m^{a}(x)\coloneqq \E[Y|X=x,A=a]$; similarly, $\gamma^a\coloneqq \Pb(Y =0|Y^*=1, A=a)$. We formulate the following assumption.
\begin{assumption}\label{ass:weak_sep_race}
$\sup_x m^{*a}(x)=1\;\; \forall a\in\{b,w\}$.
\end{assumption}
The unconditional version of assumption \ref{ass:weak_sep_race} is clearly assumption \ref{ass:weak_separability}. The following proposition can be interpreted as a generalization of proposition \ref{prop:noise_nar_gamma}.
\begin{proposition}\label{prop:estimation_noise}
Under assumptions \ref{ass:noise_process}, \ref{ass:label_dep_noise}, and \ref{ass:weak_sep_race}, the following equality holds. 
\begin{equation}
    \gamma^a=1-\sup_x m^a(x) \;\;\forall a\in\{b,w\}.
\end{equation}
\end{proposition}
Again, the convergence rate of the estimator of $\gamma^a$ is identical to the one of the estimator of $m^a(x)$.

If race-specific classifiers are trained, then this framework inherits all the results from the label-dependent noise literature. Instead, if a unique classifier is trained, with race included in the feature set, then some of the results for model training and labels correction can be adapted to this setting.

We now estimate the values of $\gamma^a$ on COMPAS data considering the setting of assumptions \ref{ass:noise_process}, \ref{ass:label_dep_noise}, and \ref{ass:weak_sep_race}. We fit one classifier for each race group and tune the parameters via cross-validation on the training set. We use extreme gradient boosted trees (xgboost) \citep{chen2016xgboost}, logistic regression (glmnet), k-nearest neighbors (knn), and support vector machines (svm). 
The resulting scores are thresholded at $1/2$ according to Bayes decision rule and the accuracy on the test set is approximately 66\% for all models and both races. The results of the estimation for estimators 1 and 2, with corresponding standard deviations, are reported in Table~\ref{tab:estimates_gamma}. 
Not surprisingly, the noise parameter for the white population is higher than that for the black population across all models. This result is a consequence of violation of the assumptions -- that are unlikely to hold in practice -- and poor performance of the models.

\begin{table}[h] 
\begin{center}
\begin{tabular}{ |c|c|c|c|c| }
\hline
Method & xgboost &  glmnet & knn & svm\\
 \hline
White/est (est 2) & 0.13 (0.11) & 0.18 (0.08) & 0.12 (0.10) & 0.15 (0.11)\\
 \hline
White/est (est 1) & 0.55 (0.02) & 0.54 (0.02) & 0.55 (0.01) & 0.54 (0.01)\\
\hline
Black/est (est 2) & 0.07 (0.06) & 0.08 (0.05) & 0.12 (0.08) & 0.10 (0.08)\\
 \hline
Black/est (est 1) & 0.44 (0.02) & 0.42 (0.02) & 0.43 (0.02) & 0.42 (0.02)\\
\hline
\end{tabular}
\caption{The mean (standard deviation) values of $\gamma^a$ estimated on 20 random train-test splits of COMPAS data are reported in the table for estimators (est) 1 and 2. The parameters of the models are tuned via cross validation. The feature set includes age, sex, count of juvenile felonies, count of juvenile misconduct, count of other juvenile charges, count of prior charges. \label{tab:estimates_gamma}}
\end{center}
\end{table}

\newpage

\section{Extension for Section \ref{sec:evaluation}}\label{sec:extension_sec_3}

\subsection{Omitted proofs} 
\label{appendix:sensitivity_proofs}
\subsubsection{Error rates and predictive parity}
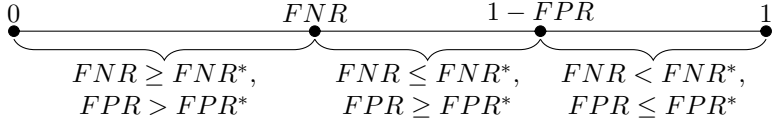
\begin{figure}[h]
\centering
\begin{tikzpicture}
\filldraw
(0,0) circle (2pt) node[align=center,   above] {0} --
(4,0) circle (2pt) node[align=center, above] {$FNR$}     -- 
(7,0) circle (2pt) node[align=right,  above] {$1-FPR$}    -- 
(10,0) circle (2pt) node[align=right,  above] {1};
\draw[decoration={brace,mirror,raise=3pt,amplitude=8pt},decorate]
  (0,0) -- node[below=9pt, align=center] {$FNR\geq FNR^*$,\\ $FPR> FPR^*$} (4,0);
\draw[decoration={brace,mirror,raise=3pt,amplitude=8pt},decorate]
  (4,0) -- node[below=9pt, align=center] {$FNR\leq FNR^*$,\\ $FPR\geq FPR^*$} (7,0); 
\draw[decoration={brace,mirror,raise=3pt,amplitude=8pt},decorate]
  (7,0) -- node[below=9pt, align=center] {$FNR < FNR^*$,\\ $FPR\leq FPR^*$} (10,0); 
\end{tikzpicture}
\caption{Possible relationships between the true and observed error rates in the case of $1 - FPR > FNR$ as described in proposition~\ref{prop:agnostic_bounds}.}
\label{fig:bounds_tikz}
\end{figure}

\textit{Proof of proposition \ref{prop:agnostic_bounds}.} Assume that $1-FPR< FNR$. We now show by contradiction that $FNR \leq FNR^*$ and $FPR\geq FPR^*$ can not hold together. Indeed, the following two equivalences hold
\begin{gather*}
    FNR \leq FNR^* \iff FNR\leq \alpha_0/\alpha\\
    FPR\geq FPR^* \iff 1-FPR\geq \alpha_0/\alpha
\end{gather*}
thanks to corollary \ref{prop:fpr_fnr_unknown_bounds}. It follows that $\alpha_0/\alpha\geq FNR> 1-FPR\geq \alpha_0/\alpha$, which is a contradiction.\\
The proof for the other case is analogous.  \hfill \qed \\
Figure~\ref{fig:bounds_tikz} provides a visual interpretation of the result.

\textit{Proof of theorem~\ref{prop:fpr_fnr_unknown_bounds}.} Recall the following notation: $p_{ij}\coloneqq \Pb(Y=i, \hat Y=j)$.
\begin{itemize}
\item {\it Proof of inequality \eqref{eq:fpr_bound}.} $FPR = \E[\hat{Y}|Y =0] $ can be rewritten as 
\begin{gather*}
\E[\hat Y|Y^*=0]\Pb(Y^*=0| Y=0)+\E[\hat{Y}|Y=0,Y^*=1]\Pb(Y^*=1| Y=0)\\
=FPR^*\;(1-\E[Y^*| Y=0])+\frac{\alpha_{1}}{\alpha} \E[Y^*| Y=0]
\end{gather*}
thanks to the law of total probability, Bayes theorem and assumption \ref{ass:noise_process} in sequence. Therefore $FPR$ is a convex combination of $FPR^*$ and $\alpha_{1}/\alpha$. Rearranging the terms we obtain
\begin{gather*}
    FPR^*=\frac{FPR-\frac{\alpha_{1}}{\alpha} \E[Y^*|Y=0]}{1-\E[Y^*|Y=0]}=\frac{p_{01}-\alpha_1}{p_{00}+p_{01}-\alpha}
\end{gather*}
For fixed $\alpha\leq \min\{p_{00},p_{01}\}$, we obtain
\begin{gather*}
    \frac{p_{01}-\alpha}{p_{00}+p_{01}-\alpha}\leq FPR^*\leq \frac{p_{01}}{p_{00}+p_{01}-\alpha}.
\end{gather*}
\item {\it Proof of inequality \eqref{eq:fnr_bound}.}  $FNR^*=\Pb(\hat{Y}=0|Y^*=1)$ can be rewritten as
\begin{gather*}
\Pb(\hat{Y}=0|Y=1)\Pb( Y=1|Y^*=1)+\Pb(\hat{Y}=0|Y=0,Y^*=1)\Pb(Y=0|Y^*=1).
\end{gather*}
Then we have
\begin{gather*}
FNR^*=FNR\; \E[Y|Y^*=1]+\frac{\alpha_{0}}{\alpha}(1-\E[ Y|Y^*=1])
\end{gather*}
which is derived as above. The last derivation follows the same strategy as above.
\item {\it Proof of inequality \eqref{eq:ppv_bound}.}  $PPV^*=\E[Y^*|\hat{Y}=1]$ can be rewritten as 
\begin{gather*}
    PPV+\E[Y^*(1- Y)|\hat{Y}=1]=PPV+\frac{\alpha_1}{p_{01}+p_{11}}.
\end{gather*}
Then, since the second term on the RHS is larger or equal to zero, the lower and upper bounds for $PPV^*$ will be given by $\alpha_1=0$ and $\alpha_1=\alpha$ respectively.
\hfill \qed\\
\end{itemize}
\textit{Proof of corollary \eqref{prop:fpr_fnr_unknown_bounds}.} Let us first prove equivalence~\eqref{eq:condition_fnr}.
\begin{gather*}
    FNR \geq FNR^* \iff \frac{p_{10}}{p_{10}+p_{11}} \geq \frac{p_{10}+\alpha_0}{p_{10}+p_{11}+\alpha} \iff FNR \geq \frac{\alpha_0}{\alpha}.
\end{gather*}
The proof of equivalence \eqref{eq:condition_fpr} for $FPR$ is similar.
\begin{gather*}
    FPR \leq FPR^* \iff \frac{p_{01}}{p_{00}+p_{01}}\leq \frac{p_{01}-\alpha_{1}}{p_{00}+p_{01}-\alpha} \iff FPR \geq \frac{\alpha_1}{\alpha}.
\end{gather*}
\hfill \qed\\
\textit{Derivation of \eqref{eq:or_fnr}.} Let us start with the case of $FNR$. If the condition in \eqref{eq:condition_fnr} holds, then we have
\begin{gather*}
\frac{p_{10}}{p_{10}+p_{11}}\geq \frac{\alpha_0}{\alpha_0+\alpha_1}\iff \alpha_1 p_{10}\geq \alpha_0 p_{11}\iff \frac{\alpha_1/\alpha_0}{p_{11}/p_{10}}\geq 1
\end{gather*}
where we used Bayes theorem and law of total probability in sequence.\\ 
The odds ratio for $FPR$ can be derived in a similar manner. For the equivalence in \eqref{eq:condition_fpr} to hold we need
\begin{gather*}
\frac{p_{01}}{p_{00}+p_{01}} \geq \frac{\alpha_1}{\alpha_0+\alpha_1}\iff \alpha_0 p_{01}\geq \alpha_1 p_{00} \iff \frac{\alpha_0/\alpha_1}{p_{00}/p_{01}}\geq 1
\end{gather*}
where we used, again, Bayes theorem and law of total probability.
\hfill \qed\\

\subsubsection{Accuracy Equity}
\textit{Proof of proposition ~\ref{prop:bounds_auc}.} The Mann-Whitney U statistic can be computed according to 
\begin{gather*}
U_1 \coloneqq R_1- \frac{n_1(n_1 + 1)}{2}=\sum_{i:Y_i=1}r_i - \frac{n_1(n_1 + 1)}{2}
\end{gather*}
where $r_i$ are the adjusted ranks. We can calculate the AUC of $S$ as a classifier of $Y^*$ from $U_1$ through the expression:
\begin{gather*}
AUC^* = \frac{U_1}{n_1n_0} = \frac{R_1}{n_1(n-n_1)} - \frac{n_1 + 1}{2(n - n_1)}
\end{gather*}
Now suppose that $\ceil{\alpha n}$ observations are unobserved recidivists. It is clear that the lower (upper) bound can be found by assuming $\ceil{\alpha n}$ observations corresponding to the lowest (highest) ranks such that $Y=0$ to be recidivists; this provides the sharp bound in the proposition. This is in turn lower (upper) bounded by the case where the lowest (highest) $\ceil{\alpha n}$ ranks overall correspond to unobserved recidivists: for the lower bound, $R_1=R_1+1+\dots+\alpha n=R_1+(\alpha n+1)\alpha n/2$, while for the upper bound, $R_1=R_1+(n-\alpha n+1)+\dots+n= R_1+\alpha n(2n-\alpha n+1)/2$. \hfill \qed \\

\subsubsection{Calibration via logistic regression}
\textit{Proof of proposition~\ref{prop:logistic_bound}.}
For a fixed set proportion of hidden recidivists $\alpha$, we aim to prove that the bounds for the coefficient of race are achieved in the settings $\alpha_1=\alpha$ and $\alpha_0=\alpha$. \\
Consider the random variables $\mathbf{X_i}=(X_{i,1},X_{i,2},X_{i,3})^T$ where $X_{i,1}=1$, $X_{i,2}\in\mathbb{R}_{+}$, and $X_{i,3}=\mathds{1}_{w}(A_i)$ with $A_i\in\{b,w\}\; \forall i\in P=\{1,\dots,n\}$. Let $W:=\{i| i\in P \text{ and } x_{i,3}=1\}$. Consider the $n$ observations $\{(y_i,\mathbf{x}_i)\}_{i=1}^n$ such that $x_{i,2}\leq x_{j,2}$ for $ 1\leq i\leq j\leq n$, that is the observations are ordered increasingly according to the realizations of $X_{i,2}$. Let $\bbeta^{\dagger}$ be the MLE of the log-likelihood \begin{gather}\label{eq:log_lkl_reg}\ell (\bbeta|\mathbf{y})=\sum_{i=1}^{n} y_i \log \sigma_{\bbeta} (\mathbf{x}_i)+(1-y_i)\log (1-\sigma_{\bbeta} (\mathbf{x}_i))\end{gather} where 
\[ \sigma_{\bbeta}(\mathbf{x})\coloneqq \frac{1}{1+e^{-\bbeta^T\mathbf{x}}}.\]
Logistic regression aims at minimizing the negative log-likelihood in \eqref{eq:log_lkl_reg}.\\
Consider two indices $l,h \in W$, $h>l$, such that $y_{l}=1$ but $y_{h}=0$.
Now let $\{(y_i^*,\mathbf{x}_i)\}_{i=1}^n$ be such that $y_i^*=y_i \; \forall i\in P\setminus \{l,h\}$; $y_{l}^*=0$ and $y_{h}^*=1$.
We are interested in the  MLE $\bbeta^{*\dagger}$ for $\ell (\bbeta|\mathbf{y}^*)$.
Consider a second-order Taylor expansion of $\ell(\bbeta|\mathbf{y}^*)$ around $\bbeta^{\dagger}$:
\begin{gather*}
    \ell(\bbeta|\mathbf{y}^*)\approx \ell(\bbeta^{\dagger}|\mathbf{y}^*)+(\bbeta-\bbeta^{\dagger})\nabla \ell(\bbeta^{\dagger}|\mathbf{y}^*)
    +\frac{1}{2}(\bbeta-\bbeta^{\dagger})^{T}\nabla^{2}\ell(\bbeta^{\dagger}|\mathbf{y}^*)(\bbeta^{*'}-\bbeta^{*}).
\end{gather*}
Note that $\nabla \ell(\bbeta|\mathbf{y}^*)|_{\bbeta=\bbeta^\dagger}=(0,x_{h,2}-x_{l,2},0)^{T}$ since $\ell (\bbeta |\mathbf{y}^*)$ can be rewritten as
\begin{gather*}
\ell(\bbeta|\mathbf{y}^*)=\ell(\bbeta|\mathbf{y})+\log \frac{\sigma_{\bbeta}(\mathbf{x}_h)}{1-\sigma_{\bbeta}(\mathbf{x}_{h})}-\log \frac{\sigma_{\bbeta}(\mathbf{x}_{l})}{1-\sigma_{\bbeta}(\mathbf{x}_{l})}
\end{gather*}
 where
\begin{gather*}
    \log \frac{\sigma_{\bbeta}(\mathbf{x})}{1-\sigma_{\bbeta}(\mathbf{x})}
    =\log \exp\{\bbeta^{T}\mathbf{x}\}=\bbeta^T \mathbf{x}
\end{gather*}
and thanks to the fact that the score evaluated at the MLE is zero. If we consider the problem of minimizing the negative log-likelihood, the Hessian is positive definite, and consequently its determinant is positive. We are interested in the direction of the search for $\bbeta^{*\dagger}$. The minimizer of the Taylor expansion above for the negative log-likelihood with respect to $\{(y_i^*, \mathbf{x}_i)\}_{i=1}^n$ is $\bbeta^{+}=\bbeta^\dagger-\left[\nabla^2\left(-\ell(\bbeta^\dagger|\mathbf{y}^*)\right)\right]^{-1}\nabla(-\ell(\bbeta^\dagger|\mathbf{y}^*)$). The Hessian is given by $\mathbf{X}^T \mathbf{D}\mathbf{X}$ where $\mathbf{D}_{ii}=s_i:=\sigma_{\bbeta}(\mathbf{x}_i)(1-\sigma_{\bbeta}(\mathbf{x}_i))$ for $i=1,\dots,n$. Therefore we have
\begin{gather*}
    \nabla^2\left(-\ell(\bbeta^\dagger|\mathbf{y}^*)\right)=\mathbf{X}^T \mathbf{D}\mathbf{X}=\begin{bmatrix}
    \sum_{i\in P}s_i       & \sum_{i\in P}s_{i}x_{i,2} & \sum_{i\in W}s_i \\
   \sum_{i\in P}x_{i,2}s_i      & \sum_{i\in P}s_i x_{i,2}^2  & \sum_{i\in W}s_i x_{i,2} \\
    \sum_{i\in W}s_i     & \sum_{i\in W}x_{i,2}s_i & \sum_{i\in W}s_i 
\end{bmatrix}
\end{gather*}
Since the gradient of $-\ell(\bbeta^\dagger|\mathbf{y}^*)$ is $(0,x_{l,2}-x_{h,2},0)$, we are only interested in the second column of the inverse of the Hessian. Through some algebra to invert the Hessian, we obtain that $\beta^{+}_{k}\leq \beta^{\dagger}_{k}$ for $k=1$ and $\beta^{+}_{k}\geq \beta^{\dagger}_{k}$ for $k=2,3$ if the following respective conditions hold:
\begin{enumerate}
    \item $\sum_{i\in P\setminus W}s_{i}x_{i,2}\geq 0$ \text{ for }k=1;
    \item $\sum_{i\in P\setminus W}s_{i}\geq 0$ \text{ for }k=2;
   \item $(\sum_{i\in W}s_{i})(\sum_{i\in P}x_{i,2})-(\sum_{i\in W}s_{i}x_{i,2})(\sum_{i\in P}s_{i})\geq 0$ \text{ for }k=3;
\end{enumerate}
where $s_{i}=\sigma_{\bbeta}(\mathbf{x}_{i})(1-\sigma_{\bbeta}(\mathbf{x}_{i}))$. Notice that condition $(2)$ will always be verified, and condition $(1)$ as well if $X_{2}\in\mathbb{R}_{+}$, as in our case. Condition $(3)$ needs to be verified case by case. It follows that, if condition $(3)$ holds for any choice of $h>l$, then the coefficient of race is a nondecreasing function of the index. \hfill \qed

For varying $\alpha$, one can prove the inequality using a similar approach. For a model with $X=(X_{1},X_{2})$ the proof is straightforward using a first-order Taylor expansion. With the inclusion of an additional covariate $X_4$, the gradient becomes $(0,x_{h,2}-x_{l,2},0,x_{h,4}-x_{l,4})^{T}$ and the inversion of the Hessian is not straightforward. 

\subsubsection{Optimization for sensitivity analysis of chi-squared conditional independence test} \label{appendix:chisq}

We recall the test statistic of the chi-squared test;
\begin{equation}
	T(h) = \sum_{k = 1}^{|S|} \sum_{\stackrel{a\in\{b,w\}}{y\in\{0,1\}}} 
    \frac{\left(O_{ay}^{(k)} - E_{ay}^{(k)}\right)^2}{E_{ay}^{(k)}},  
    \label{eq:chisqtest}
\end{equation}
The statistic is a function of the hidden recidivist counts $h = (h_1, \ldots, h_{|S|})$. Expected counts are estimated from the data assuming the null hypothesis $Y^* \ind A \mid S$ is true.  These quantities evaluate to
\begin{align*}
O_{ay}^{(k)} &= \nk_{ay} + h_k\Ind_{a=w}(2y-1), \text{ and } \\
E_{ay}^{(k)} &= \left(\nk_{wy}+\nk_{by}+ (2y-1)h_k\right)\left(\nk_{a0}+\nk_{a1}\right)/\nk.
\end{align*}
The key observation is that, when viewed as a function of $h_k$, the numerator terms $(O_{ay}^{(k)} - E_{ay}^{(k)})^2$ are convex quadratics in $h_k$, and the denominator terms $E_{ay}^{(k)}$ are linear functions in $h_k$ that are constrained to be positive.  Thus each inner summand of equation~\eqref{eq:chisqtest} is a quadratic-over-linear function, which is strongly convex \citep{boyd2004convex}.  Furthermore, since the sum of strongly convex functions is strongly convex, we can conclude that the test statistic $T$ as a function of $h$ has the form
\begin{equation}
T(h) = \sum_{k = 1}^{|S|} f_k(h_k),
\end{equation}
where each $f_k$ is a strongly convex function. This observation is important in our discussion of optimizing the test statistic subject to constraints on the hidden recidivist population.

Now, we want to maximize the test statistic \eqref{eq:chisqtest}
over $h_k$, subject to $\sum_k h_k \le N_h$.'
Note that each term is strongly convex in $h_k$, so the optimum over $h_k$ for $0 \le h_k \le C$ will always be achieved at either
$h_k=0$ or $h_k = \min\{C, \nk_{w0}\}$.  Because the objective is separable, we just take these terms in order of decreasing value in a simple greedy search:

\begin{algorithmic}
\Require {$N_h$} \Comment{Move limit}
\Require {$T_k(h_k)$} \Comment{Terms of (13) corresponding to k}
\Require {$n_{w0}^{(k)}$} \Comment{See Section \ref{sec:chisq}}
\State $B \gets N_h$
\State $h_k \gets 0$, $k=1,\dots,K$
\While {$B > 0$}
  \For{$k\gets 1$ to $K$}
     \State $r[k] \gets \max(0,T_k(\min(B, n_{w0}^{(n)} - h_k))-T_k(0))$
  \EndFor
  \If {$\max(r) \le 0$} 
  \State Break while loop
  \EndIf
  \State $i \gets \argmax(r)$ \Comment{Select greatest improvement}
  \State $h_k \gets \min(B, n_{w0}^{(n)})$
  \State $B \gets B - \min(B, n_{w0}^{(n)})$
\EndWhile
\end{algorithmic}

\subsection{Extension to one-sided label-dependent noise}
Recall from \textsection\ref{sec:estimators_nar_noise} that $\rho:=\E[Y^*|Y=0]$.

\subsubsection{Error rate balance and predictive parity.}
The following result can be read as a corollary of theorem~\ref{prop:fpr_fnr_unknown_bounds}. The decompositions of $FPR^*$ and $FNR^*$ have already been derived in \citep{jain2017recovering, scott2013classification, menon2015learning}.
\begin{corollary}\label{cor:fpr_fnr_ppv_nar} Under assumptions \ref{ass:noise_process} and \ref{ass:label_dep_noise},
\begin{gather} 
    FPR^*=\frac{FPR-\rho (1-FNR^*)}{1-\rho} \label{eq:fpr_bound_label}\\
    FNR^*=FNR \label{eq:fnr_bound_label}\\
    PPV^*=\frac{PPV}{1-\gamma} \label{eq:ppv_bound_label}
\end{gather}
\end{corollary}
{\it Proof of corollary \ref{cor:fpr_fnr_ppv_nar}.} 
\begin{itemize}
    \item {\it Proof of equation \eqref{eq:fpr_bound_label}}. Consider the decomposition \[FPR=(1-\rho)\E[\hat{Y}|Y^*=0,Y=0]+\rho \E[\hat{Y}|Y^*=1, Y=0]\] derived in the proof of theorem \ref{prop:fpr_fnr_ppv_bounds}. Then, \[\E[\hat{Y}|Y^*=0, Y=0]=\E[\hat{Y}|Y^*=0]=FPR^*\] and  \[\E[\hat{Y}|Y^*=1, Y=0]=\E[\hat{Y}|Y^*=1]:=1-FNR^*.\] The result follows.
    \item {\it Proof of equation \eqref{eq:fnr_bound_label}}. For $FNR^*$ we have
    \begin{gather*}1-FNR=\E[\hat{Y}|Y=1]=\sum_{y=0}^{1}\Pb(\hat{Y}=1,Y^*=y|Y=1)\\
    =\Pb(\hat{Y}=1,Y^*=1|Y=1)=\E[\hat{Y}|Y^*=1,Y=1]\\
    =\E[\hat{Y}|Y^*=1]=1-FNR^* \end{gather*} and therefore $FNR=FNR^*$.
    \item {\it Proof of equation \eqref{eq:ppv_bound_label}}. For $PPV^*$, similarly to the previous proofs,
    \[PPV=\E[Y|Y^*=1,\hat{Y}=1]\E[Y^*=1|\hat{Y}=1]=(1-\gamma)PPV^*.\]\hfill \qed
\end{itemize}

\subsubsection{Accuracy Equity} \begin{proposition}\label{prop:auc_nar}
Under assumptions \ref{ass:noise_process} and \ref{ass:label_dep_noise},
    \begin{equation}AUC=(1-\rho)AUC^*+\rho/2.\end{equation}
\end{proposition}
{\it Proof of proposition \ref{prop:auc_nar}}.
The resut follows from corollary 3 in \citep{menon2015learning} for the two-sided label-dependent noise setting
considering $\beta=\rho$ and $\alpha=0$. \hfill \qed

\subsubsection{Calibration via logistic regression.}\label{sec:calibration_label_dep}
Thanks to assumptions \ref{ass:noise_process} and \ref{ass:label_dep_noise}, we have
\begin{gather}
    \E[Y|S=s,A=w]=(1-\gamma)\E[Y^*|S=s,A=w]
\end{gather}
for all values of $s$, hence calibration properties can be easily checked.\\
In this context, the sensitivity analysis for calibration described in the paper (\textsection \ref{sec:evaluation}) still applies. The ``extreme'' settings $\alpha_0=\alpha$ and $\alpha_1=\alpha$ are ruled out only in expectation. In fact, when selecting hidden recidivists from the pool $\{Y=0\}$ under condition \ref{ass:label_dep_noise}, 
\[
\#\{Y\neq Y^*, \hat{Y}=0\}\sim \text{Hypergeometric}(\#\{Y=0\}, \#\{Y=0,\hat{Y}=0\}, \#\{Y\neq Y^*\}).
\]
Therefore all hidden recidivists might still happen to be in the either lowest or highest risk bins.

However, one might want to correct the model in the training phase. For this purpose, several techniques inherited from the literature on label-dependent noise can be applied.
For instance, the following two-step technique can be used: (1st step) training of {\it any} classifier and estimation of the noise rate, (2nd step) training of a logistic regression using the methods of unbiased estimators or of label-dependent costs proposed by \citep{natarajan2013learning}. We provide below a quick overview of the two methods; further details can be found in~\citep{natarajan2013learning}. 

$\cdot$ Method of unbiased estimators $\cdot$ For a scorer $f$ and a bounded loss function $\ell(f(x),y)$,
\begin{gather*}
    \E_{Y|Y^*=y^*}\left[\tilde{\ell}(f(x),Y)\right]=\ell(f(x),y^*)
\end{gather*}
where 
\begin{gather*}
    \tilde{\ell}(f(x),0)=\ell(f(x),0), \text{ and } \tilde{\ell}(f(x),1)=\frac{\ell(f(x),1)-\gamma\ell (f(x),0)}{1-\gamma}=\frac{\ell(f(x),1)(1+\gamma)-\gamma}{1-\gamma}.
\end{gather*}
The last equality is thanks to the fact that for the sigmoid loss $\ell(f(x),y)=(1+\text{e}^{-f(x)})^{y-1}(1+\text{e}^{f(x)})^{-y}$ we have $\ell(f(x),0)+\ell(f(x),1)=1$. Therefore the optimization problem on noisy labels can be solved using the loss $\tilde{\ell}$ instead of $\ell$.

$\cdot$ Method of label-dependent costs $\cdot$ For any classfier $h$, the Bayes classifier for the $0-1$ $\beta$-weighted loss function for $Y$ is
\begin{gather*}
    (1-\beta)\mathds{1}_{Y=1}\mathds{1}_{h(X)=0}+\beta\mathds{1}_{Y=0}\mathds{1}_{h(X)=1}, \qquad  \beta=(1-\gamma)/2.
\end{gather*}
This also corresponds to be the Bayes classifier for the minimization of the $0-1$ loss function for $Y$. \citep{natarajan2013learning} show that the use of the sigmoid loss $\ell$ as surrogate, that is the minimization of 
\begin{gather*}
    (1-\beta)\mathds{1}_{Y=1}\ell(f(X),1)+\beta \mathds{1}_{Y=0}\ell(f(X),0),
\end{gather*}
ensures convergence of the $0-1$ $\beta$-weighted loss function.

\newpage 
\subsection{Extension to noise in both groups} \label{app:noise_both_groups}
In this subsection we show that most of the results in our methodology extend to the case of noise in both groups without further proofs. Indeed, the results relative to error rates and AUC have been derived conditioning on the race attribute $A$. The proof for logistic regression can be easily adapted to take into account the new setting.

Let $\alpha_i^a:=\Pb(Y=0,Y^*=1,\hat{Y}=\hat{y}|A=a)$ indicate the proportion of hidden recidivists in the low ($\hat y = 0$) and high ($\hat y = 1$) risk groups for the black ($a=b$) and white ($a=b$) populations. Let $\alpha^a:=\alpha^a_0+\alpha^a_1$ be the total proportion of hidden recidivism in the population with race $a$. \\

{\bf Error rates and predictive parity.} The bounds in proposition \ref{prop:agnostic_bounds} and in theorem \ref{prop:fpr_fnr_unknown_bounds} have been obtained conditioning on the race attribute, that is $M^{*a}$ only depends on $( M^a,\alpha_0^a,\alpha_1^a)$. This means that the sensitivity analysis on the metrics of an individual race group does not depend on the noise present in other groups. Consequently the results of proposition \ref{prop:agnostic_bounds}, theorem \ref{prop:fpr_fnr_ppv_bounds}, and corollary \ref{prop:fpr_fnr_unknown_bounds} translate onto this setting without further proofs.\\
In the paper we also show that, in absence of noise for the black population, $FNR^w-FNR^{*b}\geq FNR^{*w}-FNR^{*b}$ whenever $FNR^w\geq \alpha_0^w/\alpha^w$ thanks to corollary \ref{prop:fpr_fnr_unknown_bounds}. It is clear that $FNR^w-FNR^b\geq FNR^{*w}-FNR^{*b}$ will hold if we assume $FNR^{b} \leq \alpha_0^b/\alpha^b$ and $FNR^w \geq \alpha_0^{w} /\alpha^w$; however, it is unlikely -- but not impossible -- that the inequality holds in different directions for the two populations. Therefore an interesting question is what assumptions on $(\alpha_0^w,\alpha_1^w,\alpha_0^b,\alpha_1^w)$ are needed to conclude $FNR^w-FNR^b\geq FNR^{*w}-FNR^{*b}$. Through some algebra we can retrieve the following decomposition.
\begin{gather*}
   FNR^w-FNR^b = \frac{FNR^{*w}}{\E[Y|Y^*=1,A=w]}-\frac{FNR^{*b}}{\E[Y|Y^*=1,A=b]}+\frac{\alpha_0^b}{\E[ Y|A=b]}-\frac{\alpha_0^w}{\E[ Y|A=w]}.
\end{gather*}
The {\it differential policing} assumption would suggest that $\E[Y|A=w,Y^*=1]\leq \E[Y|A=b,Y^*=1]$ therefore we can lower bound the first two terms by $FPR^{*w}-FNR^{*b}$. There only remains to show that the last two terms are larger or equal to zero. However, this is not always the case. Indeed,
\begin{gather*}
    \frac{\alpha_0^b}{\E[Y|A=b]}\geq \frac{\alpha_0^w}{\E[Y|A=w]} \iff \frac{\alpha_0^b}{\alpha_0^w}\geq \frac{\E[Y|A=b]}{\E[Y|A=w]}.
\end{gather*}
In the COMPAS data we have seen that the RHS is larger than one, but we would intuitively expect that LHS to be smaller than one. Therefore we conclude that in order to make inference on the sign of $FNR^{*w}-FNR^{*b}$, explicit assumptions on the magnitude of the noise parameters need to be formulated, that is $\alpha_0^w,\alpha_1^w,\alpha_0^b,$ and $\alpha_1^w$ need to be bounded. We do not present the computations for $FPR$, but the inequality has a similar interpretation.\\
Finally, note that theorem \ref{prop:fpr_fnr_ppv_bounds} and corollary \ref{prop:fpr_fnr_unknown_bounds} can be rewritten in terms of unconditional statements, that is on the entire population. This is the typical setup in the literature when there is no specific interest in the conditional metrics.

{\bf Accuracy Equity.} As in the case of error rates, the statement in proposition \ref{prop:bounds_auc} holds conditionally on the protected attribute. Consequently no further extension is needed.\\ Again, we remark that the statement of the proposition holds also unconditionally, or, in general, conditionally on any subset of the feature space.

{\bf Calibration via logistic regression.} We provide only a high-level idea for the extension of the proof of proposition~\ref{prop:logistic_bound}. The gradient of the log-likelihood now is $\nabla\ell(\bbeta|\mathbf{y}^*)=(0,x_{h,2}-x_{l,2},x_{h,3}-x_{l,3})$ where  $x_{h,3}-x_{l,3}$ is equal to $0$ if $x_{h,3}=x_{l,3}$, that is if the hidden recidivist does not switch race. Consequently, for a fixed configuration of hidden recidivists in one population, the bounds for the coefficients will still be achieved by the hidden recidivists taking the extreme scores. Therefore one can show that, considering a pair of hidden recidivists of different races with scores either both lower or larger than the current ones, the bounds for the coefficients are achieved in the extreme settings over the entire population.
\newpage

\subsection{Further experiments for calibration via logistic regression}\label{sec:extra_exp_logistic}
\begin{figure}[h]
  \centering
    \includegraphics[width=0.5\linewidth]{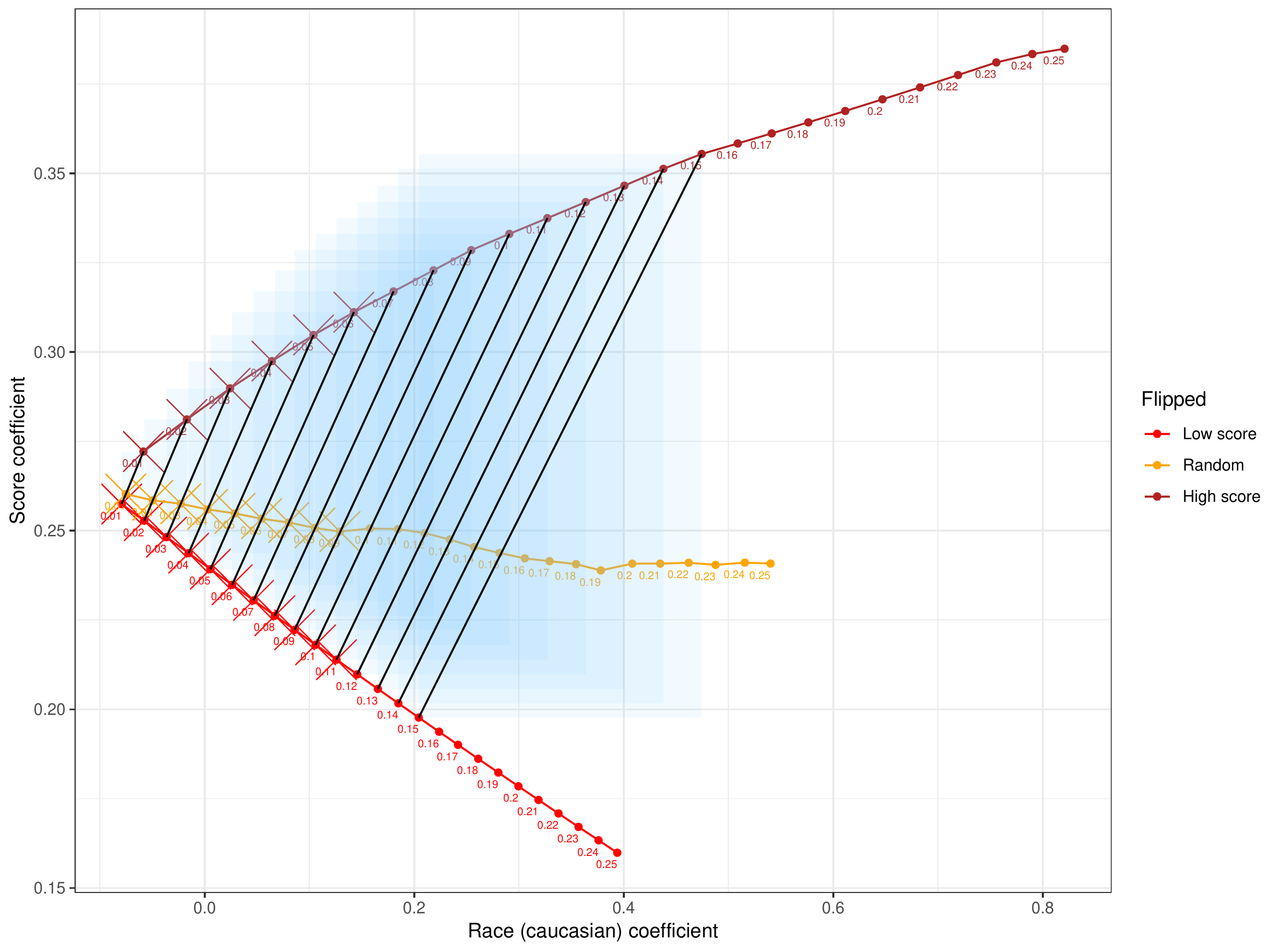}
    \caption{Outer red curves show the bounds guaranteed by proposition~\ref{prop:logistic_bound} for different choices of $\alpha$ (which are labeled on the curves). Blue rectangles correspond to analytical bounds, though empirically we can verify that the coefficients must in fact lie on the dashed diagonals.  Orange curve shows coefficients under a ``random'' mechanism where we assume each observation with $Y = 0$ was equally likely to be a hidden recidivist.  Points marked with an X correspond to values of $\alpha^w$  where the $p$-value of $\beta_w$ is \textit{not} statistically significant.  These are values of $\alpha^w$ where we would conclude that $S$ is well calibrated with respect to $A$ as a predictor of $Y^*$.}
    \label{fig:bounds_coefficients}
\end{figure}

Figure \ref{fig:bounds_coefficients} shows the two-dimensional bounds (red lines) of the coefficients of $S$ and $A$ for varying $\alpha,\alpha_0,$ and $\alpha_1$ as described by proposition \ref{prop:logistic_bound}. Although the analytical bounds for the coefficients for fixed $\alpha$ are wide, we find empirically that no matter the indexes of the hidden recidivists, the coefficients at a given $\alpha$ always lie on the diagonal (black lines) connecting the lowest and highest bounds that we find. Moreover, as previously argued, assuming label-dependent noise does not drastically change the coefficient of $S$ but only the one of $A$, as shown by the coefficients obtained ``randomly'' sampling hidden recidivists from the observations with $Y = 0$ (orange lines).

We also check calibration for race- and label-dependent noise, i.e. under assumption \ref{ass:weak_sep_race}, in the COMPAS data. The methodology follows the method of label-dependent costs described in \textsection \ref{sec:calibration_label_dep}. The procedure for the estimation of the race-specific noise rates has been described in \textsection \ref{sec:estimators_nar_noise}; we use extreme gradient boosted trees for this step~\citep{chen2016xgboost}. We resample the observations from the data set according to the weights $\beta$ described in \textsection\ref{sec:calibration_label_dep}; this is done separately within each of the two races. We then fit a logistic regression $Y\sim S+A$ on the resulting data set and check calibration via a Wald test. As in the observed data, the coefficient for $A=w$ is not statistically significant at an $\alpha$-level of $0.01$.

\newpage
\section{Error rate balance with fairness-promoting algorithms}\label{sec:fairpromote}
Through our methodology, we evaluate the effects of label noise on the error metrics of the predictions of the following four algorithms on the COMPAS data set.
We split the data into 70\% and 30\% for training and testing respectively, stratifying for race.
As feature set, we consider race, sex, age, number of juvenile felonies, misdemeanors, and other charges, count of prior arrests, degree of charge to predict two-year rearrest.
\begin{itemize}
    \item (FERM) We use the methodology proposed by \citep{donini2018empirical}, training SVM's with linear kernel to produce a classifier that approximately satisfies equal opportunity.
    \item (EQODDS) We train a logistic regression and then use the methodology described in \citep{hardt2016equality} to obtain a classifier that satisfies equal opportunity.
     \item (COMPAS6) We threshold the COMPAS decile score at 6 (i.e. $\hat{Y}=\mathds{1}(S>6)$), instead of 4.
    \item (UNCON) We train a logistic regression. 
\end{itemize}
We chose the thresholds for (COMPAS6) and logistic regression models such that the proportion of defendants predicted to be high risk was equal across all methods, i.e. around $30\%$.

\begin{figure*}[h]
\centering
  \begin{subfigure}[t]{0.47\textwidth}
   \includegraphics[width=\linewidth]{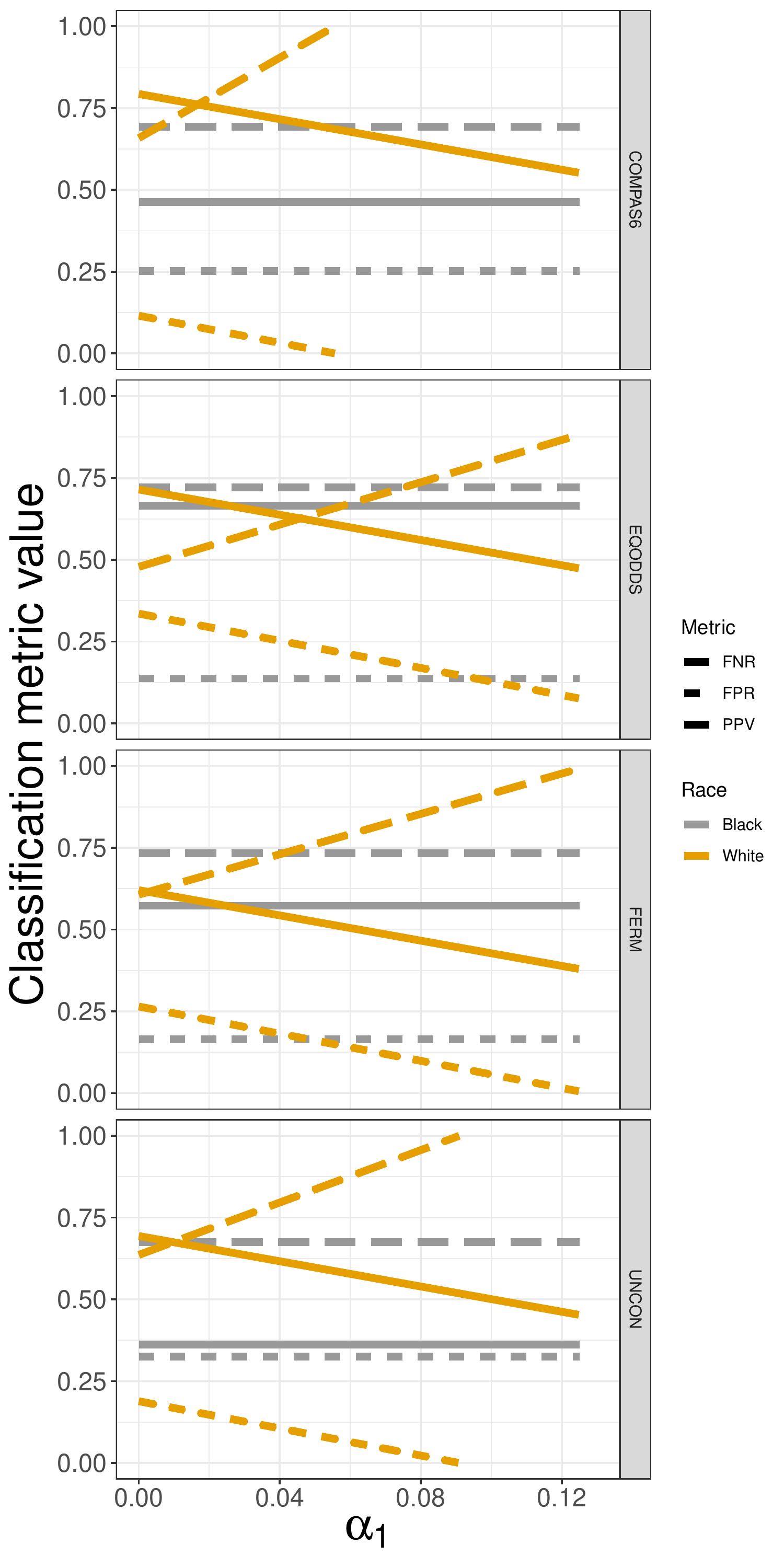}
   \caption{$\alpha=0.12$ to equalize reoffense rates across groups.}
  \end{subfigure} \hspace{2em}
  \begin{subfigure}[t]{0.47\textwidth}
   \includegraphics[width=\linewidth]{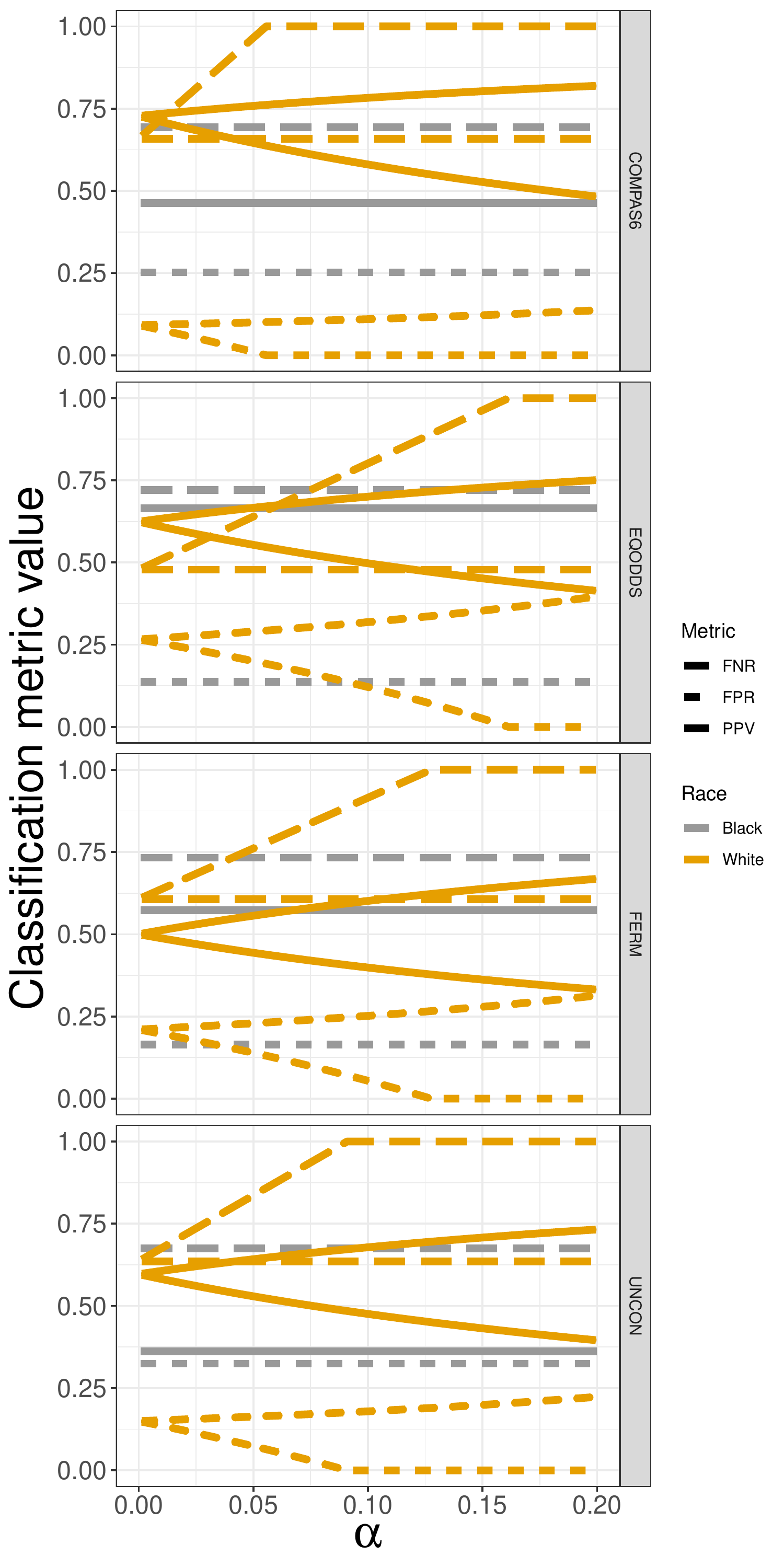}
 \caption{Bounds as described in Theorem \ref{prop:fpr_fnr_ppv_bounds} in terms of $\alpha$.}
  \end{subfigure} \hspace{2em}
\caption{Analysis of predictive parity and error rate balance for COMPAS across different TVB scenarios for four different algorithms, as described in the text.
Orange lines show values of $FPR^w$, $FNR^w$, and $PPV^w$.  Grey lines show corresponding values for the black population. }\label{fig:bouns_debias}
\end{figure*}

The bounds for the error metrics for the predictions of the four classifiers as functions of the noise are shown in Figure~\ref{fig:bouns_debias}. 
For varying $\alpha$, we observe that the classifiers in (COMPAS6) and (UNCON) do not satisfy error rate balance on the observed labels. Due to the large differences in error rates, equality of the metrics of these models cannot be achieved by any configuration of the noise for $\alpha\leq 0.2$. Differently, equality is possible for (EQODDS) and (FERM) for $\alpha$ larger than $0.08$, well below the level of $0.12$ necessary to equalize reoffense rates.
When the noise is fixed at $\alpha=0.12$, we observe a similar pattern. Despite the unavoidable degree of uncertainty, (FERM) comes close to achieving parity: for $\alpha_1=0.04$, the metrics of (FERM) are approximately equal across populations. These results suggest that the two presented fairness-promoting methods perform better than unconstrained methods under label noise.

\end{document}